\documentclass[letterpaper]{JHEP3}
\usepackage{epsfig}
\usepackage{comment}

\newcommand{\roughly}[1]{\mathrel{\raise.3ex\hbox{$#1$\kern-0.85em
\lower1ex\hbox{$\sim$}}}}

\newcommand{\gsim}{\roughly>}

\def\bea{\begin{eqnarray}}
\def\eea{\end{eqnarray}}
\def\ba{\begin{eqnarray}}
\def\ea{\end{eqnarray}}
\def\be{\begin{equation}}
\def\ee{\end{equation}}

\def\ssA{{\scriptscriptstyle A}}
\def\ssB{{\scriptscriptstyle B}}
\def\ssC{{\scriptscriptstyle C}}
\def\ssD{{\scriptscriptstyle D}}

\def\ssL{{\scriptscriptstyle L}}

\def\ssY{{\scriptscriptstyle Y}}

\def\B{\mathcal{B}}
\def\C{\mathcal{C}}

\def\nn{\nonumber}

\def\({\left(}
\def\){\right)}

\def\pref#1{(\ref{#1})}

\title{Light Octet Scalars, a Heavy Higgs\\ and
Minimal Flavour Violation}

\author{C.P. Burgess,
${}^{1,2}$ Michael Trott,${}^{1}$\
and Saba Zuberi${}^{3}$ \\
${}^1$ Perimeter Institute for Theoretical Physics,\\
\hspace{2cm} 31 Caroline St. N., Waterloo ON, N2L 2Y5, Canada.\\
${}^2$ Dept. of Physics \& Astronomy, McMaster University,\\
\hspace{2cm} 1280 Main St. W., Hamilton ON, L8S 4M1, Canada.\\
${}^3$ Dept. of Physics, University of Toronto,\\
 \hspace{2cm} 60 St. George Street, Toronto ON, M5S 1A7, Canada.
}

\date{}

\abstract {It is widely believed that existing electroweak data
requires a Standard Model Higgs to be light while electroweak and
flavour physics constraints require other scalars charged under
the Standard Model gauge couplings to be heavy. We analyze the
robustness of these beliefs within a general scalar sector and
find both to be incorrect, provided that the scalar sector
approximately preserves custodial symmetry and minimal flavour
violation (MFV). We demonstrate this by considering the
phenomenology of the Standard Model supplemented by a scalar
having $\rm SU_c(3) \times SU_\ssL(2) \times U_\ssY(1)$ quantum
numbers $({\bf8}, {\bf2})_{1/2}$ --- which has been argued
\cite{Manohar:2006ga} to be the only kind of exotic flavour singlet scalar allowed
by MFV that couples to quarks. We examine constraints coming from electroweak precision data, direct production from LEPII and the Tevatron, and from
flavour physics, and find that the observations allow both the
Standard Model Higgs and the new scalars to be simultaneously
light --- with masses $\sim {\rm 100 \, GeV}$, and in some cases
lighter. The discovery of such light coloured scalars could be a
compelling possibility for early LHC runs, due to their large
production cross section, $\sigma \sim 100 \, {\rm pb}$. But the
observations equally allow all the scalars to be heavy (including
the Higgs), with masses $\sim {\rm 1 \, TeV}$, with the presence
of the new scalars removing the light-Higgs preference that
normally emerges from fits to the electroweak precision data. }

\begin{document}
\section{Introduction}

Most physicists believe that new physics beyond the Standard Model
(SM) awaits discovery at the LHC, and experiments at the Large
Hadron Collider (LHC) will soon probe the weak scale and
(hopefully) reveal the nature of whatever new physics lies beyond
the Standard Model. Since the Higgs sector is among the least
understood in the SM, new scalar physics could well be what is
found.

However, to be found at the Tevatron or the LHC, any such new
scalar physics should be associated with a comparatively low
scale, $\Lambda \sim \rm TeV$. And because the scale is low, it
must be checked that the new physics cannot contribute to
processes that are well-measured and agree well with the SM, such
as electroweak precision data (EWPD) and flavour-changing neutral
currents (FCNCs). This suggests taking most seriously those kinds
of new physics that suppress such contributions in a
natural way. This can be elegantly accomplished if the effective
field theory (EFT) appropriate to low energies obeys approximate
symmetries, such as a custodial symmetry ${\rm SU}(2)_C$
\cite{Susskind:1978ms,Weinberg:1979bn,Sikivie:1980hm} for EWPD and
the principle of minimal flavor violation (MFV)
\cite{Chivukula:1987py,
Hall:1990ac,D'Ambrosio:2002ex,Cirigliano:2005ck,Buras:2003jf,Branco:2006hz},
which suppresses FCNCs when formulated appropriately
\cite{Feldmann:2008ja,Kagan:2009bn,Feldmann:2009dc}.

Recently, it was discovered \cite{Manohar:2006ga} that there are
comparatively few kinds of exotic scalars that are flavour singlets and can have Yukawa
couplings with SM fermions in a way that is consistent with MFV.
The only two possible scalar representations allowed are those of
the SM Higgs or octet scalars, respectively transforming under the
gauge group ${\rm SU}(3) \times {\rm SU}(2) \times {\rm U}(1)$ as
$(\bf{1},\bf{2})_{1/2}$ or $(\bf{8},\bf{2})_{1/2}$.

In this paper we examine what constraints EWPD\footnote{We thank
J. Erler for private communication on the recent update to the
EWPD fit results related to \cite{Erler:2009jh}.}, flavour
physics, and direct production constraints place on this general
scalar sector consistent with MFV. To this end we consider the
Manohar-Wise model, for which only one $(\bf{1},\bf{2})_{1/2}$
scalar and one $(\bf{8},\bf{2})_{1/2}$ scalar are present.

Since it is the quality of SM fits to electroweak precision data
that at present provide our only direct evidence for the existence
of the SM Higgs, it is perhaps not surprising that the existence
of a scalar octet can alter the Higgs properties to which such
fits point. In particular, the best-fit value of the Higgs mass
obtained from SM fits to EWPD is now $96^{+ 29}_{-24} \, {\rm
GeV}$ \cite{Erler:2009jh}. We find that for the Manohar-Wise
model, EWPD fits both change the implications for the Higgs mass,
and limit the allowed mass range of the extended scalar sector.

We find that when the masses of the Higgs and octet
scalars are approximately degenerate, the electroweak fits allow
both the Higgs and the octet to be light, with masses $\sim 100 \,
{\rm GeV}$ (or even lighter for some components). Alternatively,
agreement with EWPD also allows the octet and the Higgs doublets
to be both heavy, with masses $\sim 1 \, {\rm TeV}$. The Higgs
doublet can be heavy and remain consistent with precision fits
because its contribution to the relevant observables is partially
cancelled by the contribution of the octet doublet. Having such a
heavy Higgs without ruining electroweak fits is attractive, as a
resolution of the so-called `LEP Paradox' \cite{Barbieri:2000gf}.
We find that the precision electroweak fits generically prefer to
limit the splittings among some of the octet components, but by an
amount that does not require fine tuning of parameters in the
potential. (The overall masses of the two multiplets are subject
to the usual issues associated with the electroweak hierarchy.)

The plan of this paper is as follows, in Section 2 we review the
Manohar-Wise model, and describe its motivation as a general
scalar sector that can both allow an approximate custodial
symmetry and satisfy MFV. In Section 3 we present our results for
the phenomenology of the model. In particular, we describe its
implications for an EWPD fit, and explore the parameter space that
allows both doublets to be either light or heavy. Since the fits
prefer a scalar spectrum that is approximately custodially
symmetric, we also study loop-induced ${\rm
SU}(2)_C$ breaking, and demonstrate that the allowed parameter
space is not fine tuned. This section also describes
direct-production constraints on the Higgs and octet scalar,
coming from both LEP2 and the Tevatron, and reexamines how
previously studied flavour constraints change if the new octets
are comparatively light. We find that the octets can pass all
these tests, for parameters with scalars that are either light or
heavy. Some conclusions are briefly summarized in Section 4.

\section{Theory}

In this section we recap the main features of the the model,
obtained by supplementing the SM with an colour-octet,
$SU_\ssL(2)$-doublet scalar. Particular attention is spent on its
approximate symmetries, since these underly the motivation to
naturally satisfy FCNC and EWPD constraints.

\subsubsection*{Motivation for $({\bf8},{\bf2})_{1/2}$ scalars.}

Minimal Flavour Violation (MFV) is a framework for having
flavour-dependent masses without introducing unwanted flavour
changing neutral currents (FCNCs). It assumes all breaking of the
underlying approximate ${\rm SU}(3)_U \times {\rm SU}(3)_D \times
{\rm SU}(3)_Q$ flavour symmetry of the SM is proportional to the
up- or down-quark Yukawa matrices. The fact that only
scalars transforming as $({\bf8},{\bf2})_{1/2}$, or as the SM
Higgs \cite{Manohar:2006ga}, can Yukawa couple to SM fermions
consistent with MFV is the motivation of the phenomenological
study we present here.

However, we also note that octet scalars appear in many specific
new-physics scenarios, including various SUSY constructions
\cite{Plehn:2008ae,Choi:2008ub}, topcolour models
\cite{Hill:1991at}, and models with extra dimensions
\cite{Dobrescu:2007yp,Dobrescu:2007xf}. Various approaches to
grand unification also have light colour octet scalars, including
Pati-Salam unification \cite{Popov:2005wz} and $\rm SU(5)$
unification
\cite{Dorsner:2007fy,FileviezPerez:2008ib,Perez:2008ry}.  Colour
octet doublets have also recently been used to study new
mechanisms for neutrino mass generation
\cite{FileviezPerez:2009ud}. Octet scalar doublets appear
naturally in models of the Chiral-Colour
\cite{Frampton:1987dn,Idilbi:2009cc} type where QCD originates in
the chiral colour group $\rm SU_L(3) \times SU_R(3)$, since in
this case octet doublets are expected in addition to the Higgs as
$\bf  3 \otimes \bf \bar{3} = \bf 8 \oplus \bf 1$. As discussed in
\cite{Gerbush:2007fe} one can also consider the class of models
where the SM is extended with $\rm SU(N) \times SU(3)_C \times
SU(2)_L \times U(1)_Y$ and imagine model-building composite Higgs
models with a $({\bf8},{\bf2})_{1/2}$ scalar in the low energy
spectrum. We emphasize that although many BSM scenarios contain
$({\bf8},{\bf2})_{1/2}$ scalars our motivation is essentially
phenomenological.

\subsection{The Manohar-Wise model}

In the Manohar Wise model \cite{Manohar:2006ga}, the scalar sector
of the SM is supplemented with the $(\bf{8},\bf{2})_{1/2}$ scalar
denoted
\bea
{S}^A = \left(\begin{array}{c} {S^A}^+ \\
{S^A}^0 \end{array} \right) \eea
where $A$ is the colour index.

The Yukawa couplings of the $(\bf{8},\bf{2})_{1/2}$ scalar to
quarks is determined up to overall complex constants, $\eta_U$ and
$\eta_D$, to be
\bea
 L = \eta_U \, g_{i j}^U \, \bar{u}_{R}^i T^A (S^A)^T \, \epsilon \, Q^j_L -\eta_D \, g_{i j}^D \, \bar{d}_{R}^i T^A (S^A)^\dagger \,Q^j_L  + h.c,
\eea
where $g^U$ and $g^D$ are the standard model Yukawa matrices,
$i,j$ are flavor indices and
\bea \label{eps}
\epsilon = \left(\begin{array}{cc} 0 & 1 \\
-1 & 0 \end{array}  \right).
\eea
 The most general renormalizable potential \cite{Manohar:2006ga} is
\bea \label{MWpotential}
 V &=&  \frac{\lambda}{4} \left(H^{\dagger \, i} H_i
 - \frac{v^2}{2} \right)^2 + 2 m_S^2 \, {\rm Tr}\, \left(
 S^{\dagger \, i} S_i \right)
 + \lambda_1 H^{\dagger \, i}H_i \, {\rm Tr}\,
 \left( S^{\dagger \, j} S_j \right)
 + \lambda_2 H^{\dagger \, i}H_j \, {\rm Tr}\,
 \left( S^{\dagger \, j} S_i \right) \nn \\
 &+& \left[\lambda_3 H^{\dagger \, i} H^{\dagger \, j} \,
 {\rm Tr}\, \left( S_i S_j \right)
 + \lambda_4 H^{\dagger \, i} \, {\rm Tr}\, \left(
 S^{\dagger \, j} S_j S_i \right) +\lambda_5 H^{\dagger \, i}
 \, {\rm Tr}\, \left( S^{\dagger \, j} S_i S_j \right)
 + h.c.\right] \nn \\
 &+&\lambda_6 {\rm Tr}\, \left( S^{\dagger \, i} S_i
 S^{\dagger \, j} S_j \right)
 + \lambda_7 {\rm Tr} \, \left(  S^{\dagger \, i} S_j
 S^{\dagger \, j} S_i \right)
 + \lambda_8 {\rm Tr}\,\left( S^{\dagger \, i} S_i\right)
 \, {\rm Tr} \, \left( S^{\dagger \, j} S_j  \right) \nn \\
 &+& \lambda_9 {\rm Tr}\, \left( S^{\dagger \, i} S_j
 \right) \, {\rm Tr} \, \left( S^{\dagger \, j} S_i \right)
 + \lambda_{10} {\rm Tr} \, \left( S_i S_j \right) \,
 {\rm Tr} \, \left( S^{\dagger \, i} S^{\dagger \, j} \right)
 + \lambda_{11} {\rm Tr}\, \left( S_i S_j \,S^{\dagger \, j}
 S^{\dagger \, i} \right) \,,
\eea
where $i$ and $j$ are ${\rm SU}(2)$ indices and $S = S^A \, T^A$.
Since a field redefinition can be used to make $\lambda_3$ real,
this represents $14$ real parameters in the potential beyond those
of the SM, which reduce to $9$ in the custodial $\rm SU(2)$
symmetric case --- see eqs.~\pref{lambdacust1} through
\pref{lambdacust2}, below. No new parameters enter in the
couplings of the $(\bf{8},\bf{2})_{1/2}$ scalar to the electroweak
gauge bosons since it has the same electroweak quantum numbers as
the Higgs. We use this fact to bound the masses of the octets in
Section \ref{EWPDcons}. The $\lambda_{1,2,3}$ terms in
Eq.(\ref{MWpotential}) lift the mass degeneracy of the octet
states when the Higgs acquires a vacuum expectation value.
Expanding the neutral scalar octet as
\bea
S^{A\,0} = \frac{S^{A\,0} _R + i S^{A\,0}_I}{\sqrt{2}}
\eea
the tree level masses become \cite{Manohar:2006ga}
\bea \label{treemass}
M_{\pm}^2 &=& M_S^2+ \lambda_1 \frac{v^2}{4} \nn \\
M_{R}^2 &=&M_S^2+ \left(\lambda_1+ \lambda_2+2 \lambda_3 \right) \frac{v^2}{4} \nn \\
M_{I}^2 &=&M_S^2+ \left(\lambda_1+ \lambda_2-2 \lambda_3 \right) \frac{v^2}{4}.
\eea

\subsubsection{Custodial symmetry}\label{cust}

We find below that EWPD fits prefer the masses of some of the
scalars in these models to be approximately degenerate in mass. In
particular, fits prefer a mass pattern that can be naturally
understood as being due to an approximate custodial $\rm SU(2)_C$
symmetry, under which the SM vector bosons transform as a triplet
and the Higgs transforms as a singlet and a triplet. This symmetry
is broken in the SM both by hypercharge gauge interactions, and by
the mass splittings within fermion electroweak doublets.

For these reasons we next explore the implications of the
custodial-invariant limit, for which $\rm SU(2)_C$ is an exact
symmetry of the underlying new physics beyond the SM. In this
scenario, it is interesting to examine the case that $\rm SU(2)_C$
is preserved in the Manohar-Wise model potential at a high scale
$\sim 1 \, {\rm TeV}$, up to the breaking that must be induced by
the SM. Imposing exact $\rm SU(2)_C$ on the octet Higgs potential
we find that the potential can be rewritten in terms of
bi-doublets
\ba \Phi = \, \left(\epsilon \, \phi^\star, \phi
\right), \quad \quad  \mathcal{S}_A =( \epsilon \, S^\star_A,
S_A), \ea
where $\epsilon$ is given in Eqn.~(\ref{eps}) and the most general
gauge- and custodial-invariant potential becomes
\ba
 V = \frac{\lambda}{16} \, \left[ {\rm Tr} \, \left(
 \Phi^\dagger \, \Phi \right)  - v^2 \right]^2
 + \frac{m_S^2}{2} \, {\rm Tr} \, \left(
 {\mathcal S_A}^\dagger  {\mathcal S_A} \right)
 + \frac{\lambda_1}{8} \, {\rm Tr} \, \left(  \Phi^\dagger
 \, \Phi \right)  \, {\rm Tr} \, \left(
 {\mathcal S_A}^\dagger  {\mathcal S_A} \right), \nn  \\
 + a_1  \, {\rm Tr} \, \left( {\mathcal S}^\dagger
 \,\Phi \right) {\rm Tr} \left( {\mathcal S}^\dagger \,
 \Phi  \right) + \left(b_1 \, {\rm Tr[T^A \,  T^B \, T^C] }
  {\rm Tr} \, \left( \Phi^\dagger  \, {\mathcal S}_A \, {\mathcal S}^\dagger_B
 \,  {\mathcal S}_C \right)  + h.c. \right)\nn \\
+  c_1 \, {\rm Tr[T^A \,  T^B \, T^C] }
 {\rm Tr} \, \left( {\mathcal S}^\dagger_A
 \, {\mathcal S_C} \right)  \,  {\rm Tr} \left(
 {\mathcal S}^\dagger_B  \, \Phi \right), \nn \\
 + d_1 \, {\rm Tr [T^A \,  T^B \, T^C \, T^D] }
 {\rm Tr} \, \left( {\mathcal S}^\dagger_A \,
 {\mathcal S_B} \right)  \,  {\rm Tr} \left(
 {\mathcal S}^\dagger_C  \,  {\mathcal S_D}\right) , \nn \\
 +e_1 \, {\rm Tr[T^A \,  T^B] \, Tr[T^C \, T^D] }
  {\rm Tr} \, \left( {\mathcal S}^\dagger_A \,
  {\mathcal S_B} \right)  \,  {\rm Tr} \left(
  {\mathcal S}^\dagger_C  \,  {\mathcal S_D} \right), \nn \\
 +f_1 \, {\rm Tr[T^A \,  T^B] \, Tr[T^C \, T^D] }
 {\rm Tr} \, \left( {\mathcal S}^\dagger_A \, {\mathcal S_C}
 \right)  \, {\rm Tr} \left( {\mathcal S}^\dagger_B
 \,  {\mathcal S_D} \right),
\ea
where $\rm T_A$ is used as a basis in colour space with 9 independent terms when the potential is $\rm
SU(2)_C$ invariant.\footnote{An alternative way to obtain this
count is to regard SU(2)${}_L \times $SU(2)${}_C$ as $SO(4)$, with
both $\vec H$ and $\vec S^\ssA$ transforming as real fields in the
4-dimensional representation. In this case the invariants of the
potential can be written $m_S^2 ({\vec S}^\ssA \cdot {\vec S}^\ssA
)$, $d_{\ssA\ssB\ssC} ({\vec H} \cdot {\vec S}^\ssA )({\vec
S}^\ssB \cdot {\vec S}^\ssC )$, $f_{\ssA\ssB\ssC} ({\vec H}_i \cdot {\vec S}_j^\ssA {\vec
S}_k^\ssB \cdot {\vec S}_l^\ssC ) \, \epsilon^{ijkl}$, $({\vec H} \cdot {\vec H}) ( {\vec
S}^\ssA \cdot {\vec S}^\ssA )$, $( {\vec H} \cdot {\vec S}^\ssA )(
{\vec H} \cdot {\vec S}^\ssA )$, $({\vec S}^\ssA \cdot {\vec
S}^\ssA )^2$ and the two independent ways of colour-contracting
$({\vec S}^\ssA \cdot {\vec S}^\ssB ) ({\vec S}^\ssC \cdot {\vec
S}^\ssD )$.} Expanding out the potential and comparing to the
general result of eq.~\pref{MWpotential}, we confirm the result of
\cite{Manohar:2006ga} that $\rm SU(2)_C$ implies
\ba \label{lambdacust1}
 2 \, \lambda_3 &=& \lambda_2, \\
 2 \, \lambda_6 &=& 2 \, \lambda_7 = \lambda_{11}, \\
 \lambda_9 &=& \lambda_{10} \,,
\ea
but we also find the additional constraint\footnote{We thank A Manohar for communication on this point clearing up a subtlety.}
%
\ba \label{lambdacust2}
 \lambda_4 = \lambda_5^\star.
\ea
Note that this constraint can effect the production
mechanism of the octets at Tevatron and LHC. We see in particular
that because $\rm \rm SU(2)_C$ symmetry implies $\lambda_2=2
\lambda_3$, in this limit $M_\pm$ and $M_I$ become degenerate.

\subsection{Naturalness issues}\label{naturalness}

In general, even if the scalar potential is required to be
custodial invariant at a particular scale, it does not remain so
under renormalization due to the presence of custodial-breaking
interactions within the SM itself. In this section we compute
these one-loop symmetry breaking effects, allowing us to quantify
the extent to which the custodial-invariant potential is
fine-tuned. To do so we calculate in Feynman gauge and note that
ghost fields do not couple to the components of the $S$ doublet.
We also neglect goldstone boson contributions to the mass
splitting as they come from the $\rm SU(2)_C$ symmetric potential
and so therefore cancel out in the mass splittings; not leading to
mixing between the $S_R$ and $S_I$ states.

\subsubsection*{$\rm SU(2)_C$  breaking due to Yukawa corrections}

\begin{figure}[hbtp]
\centerline{\scalebox{1}{\includegraphics{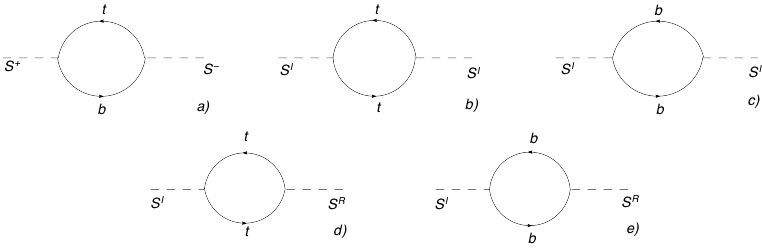}}}
\caption{$\rm SU(2)$ violating contributions to $S^I,S^{\pm}$
masses from the yukawa sector of the theory.} \label{yukawa}
\end{figure}

The breaking of  $\rm SU(2)_C$ due to Yukawa couplings is
straightforward, the requisite diagrams are given by Fig
\ref{yukawa}.

The correction to the mass $S^- \, S^+$  two point function comes
from diagram (a) and is given by
\bea
 \delta \langle T\{S^+ \, S^-\} \rangle_Y &=& -
 \delta_{ab}\, \frac{(m_b^2 \, |\eta_D|^2
 + m_t^2 \, |\eta_U|^2)[A_0(m_b^2) + A_0(m_t^2)
 - p^2 B_0(p^2,m_b^2,m_t^2)]}{16 \, \pi^2 v^2}  \\
 &\, & \hspace{-2cm} - \delta_{ab} \, \frac{(m_b^4 \, |\eta_D|^2+
 m_t^4 \, |\eta_U|^2 + m_b^2 \, m_t^2 (|\eta_D|^2 + |\eta_U|^2 - 2
 \, \eta_D \, \eta_U - 2 \, \eta^\star_D \, \eta^\star_U))\,
 B_0(p^2,m_b^2,m_t^2))}{16 \, \pi^2 v^2} \nn
\eea
where we express our results in terms of Passarino-Veltman (PV)
functions whose definitions are given in \cite{Wells:2005vk}, and
we set $|V_{tb}| \simeq 1$.

The contributions to the $S_{I}^2$ operator comes from the
diagrams (b) and (c) and is given by
\bea
 \delta \langle T\{S^I \, S^I\} \rangle_Y &=&
 - \delta_{ab} \,  \frac{m_t^2 (2 A_0(m_t^2) \,
 |\eta_U|^2  +  B_0(p^2,m_t^2,m_t^2)(4 \, m_t^2 \,
 {\rm Im}[\eta_U]^2 - p^2 \,  |\eta_U|^2))}{16 \,
 \pi^2 v^2}, \nn \\
 &-&  \delta_{ab} \, \frac{m_b^2 (2 A_0(m_b^2) \,
 |\eta_D|^2  + B_0(p^2,m_b^2,m_b^2)(4 \, m_b^2 \,
 {\rm Im}[\eta_D]^2 - p^2 \, |\eta_D|^2))}{16 \, \pi^2 v^2}.
\eea

We are interested in the mass splitting  of $M_I^2$ and $M_\pm^2$,
however to the accuracy we work one can also easily calculate the
shifts to $\delta \langle T\{S^R \, S^R\} \rangle_Y  $ and $\delta
\langle T\{S^R \, S^I\} \rangle_Y $ due to the mixing induced
between the real and imaginary components of $S^{A0}$. With these
results we can then obtain the contributions to the diagonalized
$M'_I$. The correction to $\delta \langle T\{S^R \, S^R\}
\rangle_Y $ is given by the same diagrams as $\delta \langle
T\{S^I \, S^I\} \rangle_Y $ with the appropriate replacements,
giving
\bea
 \delta \langle T\{S^R \, S^R\} \rangle_Y &=&
 -\delta_{ab} \, \frac{m_t^2 (2 A_0(m_t^2) \,
 |\eta_U|^2  +  B_0(p^2,m_t^2,m_t^2)(4 \, m_t^2
 \, {\rm Re}[\eta_U]^2 - p^2 \,  |\eta_U|^2))}{16 \,
 \pi^2 v^2}, \nn \\
 &-&  \delta_{ab} \,\frac{m_b^2 (2 A_0(m_b^2)
 \, |\eta_D|^2  + B_0(p^2,m_b^2,m_b^2)(4 \, m_b^2
 \, {\rm Re}[\eta_D]^2 - p^2 \, |\eta_D|^2))}{16 \, \pi^2 v^2}.
\eea

The  mixing of the $S_R,S_I$ fields at one loop $ \delta \langle T\{S^R \, S^I\} \rangle_Y $ is given by diagrams (d,e) and is given by
\bea
 \delta \langle T\{S^R \, S^I\} \rangle_Y&=&
 -\delta_{ab}\frac{(m_b^4 \, {\rm Re}[\eta_D] \, {\rm Im}[\eta_D]
 \, B_0(p^2,m_b^2,m_b^2) - m_t^4 \, {\rm Re}[\eta_u] \, {\rm
 Im}[\eta_U] \, B_0(p^2,m_t^2,m_t^2) )}{4 \, \pi^2 \, v^2} \nn
\eea
which is only nonzero when at least one of the MFV proportionality
constants $\eta_D,\eta_U$ are imaginary as expected. We define the
mixing angle and renormalize the theory in the Appendix.

\subsubsection*{Gauge sector $\rm SU(2)_C$ violating
corrections}

\begin{figure}[hbtp]
\centerline{\scalebox{1.2}{\includegraphics{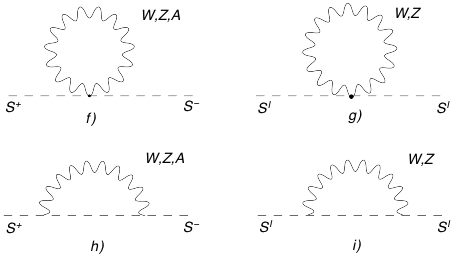}}}
\caption{$\rm SU(2)$ violating contributions from the gauge sector
of the theory.} \label{gaugequad}
\end{figure}

Calculating the required four diagrams represented by diagrams
(g,i) in Fig \ref{gaugequad} one finds
\bea
 \delta \langle T\{S^I \, S^I\} \rangle_G &=& \frac{g_1^2}{16
 \pi^2} \delta^{AB} \left( \frac{d A_0[M_W^2]}{2} +\frac{d
 A_0[M_Z^2]}{4 c_W^2} -\frac{1}{2} I_3[p^2, M_{W}^2,M_{\pm}^2]
 -\frac{1}{4 c_W^2} I_3[p^2, M_{Z}^2,M_R^2]\right) \nn
\eea
where $c_W \equiv cos[\theta_W]$ and the integral is given in
terms of PV functions as  follows
\bea
 I_3 [p^2, M_{a}^2,M_b^2] =(2 \, p^2+2 M_b^2-M_{a}^2)
 B_0[p^2,M_{a}^2,M_b^2]+ 2 A_0[M_{a}^2]-A_0[M_b^2] .
\eea

The result for $\delta \langle T\{S^R \, S^R\} \rangle_G$ is
identical up to the replacement $ M_R \rightarrow M_I$. One can
similarly calculate the other six diagrams corresponding to (f,h)
that give the following contribution for $\delta \langle T\{S^+ \,
S^-\} \rangle_G$ in terms of PV functions\footnote{Note that
diagram (f) with a photon loop is scaleless and vanishes in dim
reg.}
\bea
 \delta \, \langle T\{S^+ \, S^-\} \rangle_G &=&
 \frac{g_1^2}{16 \pi^2} \, \delta^{AB}
 \left( \frac{d A_0[M_W^2]}{2} +\frac{d (1-2 s_W^2)^2
 A_0[M_Z^2]}{4 c_W^2}-\frac{1}{4} I_3[p^2,
 M_{W}^2,M_R^2] \right. \nn \\
 &\,& \left.  \hspace{-1cm} -\frac{1}{4} I_3[p^2,
 M_{W}^2,M_I^2]-\frac{(1-2 s_W^2 )^2}{4 c_W^2} I_3[p^2,
 M_{Z}^2,M_\pm^2]- s^2 I_3[p^2, 0, M_\pm^2] \right)
\eea
Mixing between the states $S^I,S^R$ is forbidden in the gauge
sector as the couplings are real.


Given these loop-generated effects, we wish to estimate how large
the custodial-symmetry-breaking interactions are once we run down
to observable energies from the scale of new physics. The answer
depends on how far we must run, however due to the hierarchy
problem of the Higgs mass (which is only accentuated when more
light scalars are added to the spectrum), it is likely that new
physics must intervene at a relatively low scale for new physics
of $\sim {\rm TeV}$. Such a low scale for a UV completion implies
that the symmetry structure of the UV theory is consistent with
EWPD and flavour constraints.

The splitting induced by SM interactions is given by the
difference between the renormalized mass at $\Lambda$ and the low
scale, where we ignore the running for simplicity in this estimate
\bea
 \int_\Lambda^{m} (\frac{\partial \, M_i^2}{\partial \, \mu})
 \,  \partial \mu =   M_i^2 \, \left[Z^{\alpha}_{Mi}(\mu = \Lambda)
 - Z^{\alpha}_{Mi}(\mu = m)\right],
\eea
where $Z^{\alpha}_{Mi}$ is the leading perturbative correction of
the mass counterterms, whose values are given explicitly in the
appendix using a zero-momentum subtraction scheme.

As is shown in detail in the next section, the largest $M_I,
M_{\pm}$ $\rm SU(2)_C$ violating mass-splitting that is allowed by
our EWPD fit is approximately $\sim 40 (55) \, {\rm GeV}$ for the
entire $68 \% (95 \%)$ confidence regions (see Figure
\ref{figure3}). We now examine how natural such a small splitting
is assuming a typical low mass of $150 \, {\rm GeV}$.

In determining the splitting, the values of $\eta_i$ employed are
critical. For the lower bound on the $\eta_i$ we take the one
approximate loop radiatively induced value $\eta_i \sim 0.35^2/(16
\, \pi^2) $. Note that we use the result of  \cite{Gresham:2007ri}
that determined an upper bound on $|\eta_U|$ from the effect of
the octet on $R_b = (Z \rightarrow \bar{b} \, b)/(Z \rightarrow
{\rm Hadrons})$. For charged scalar masses of $(75,100,200) \,
{\rm GeV}$ the one sigma allowed upper value for $|\eta_U|$ is
$(0.27,0.28,0.33)$.

For $M_{\pm} = 150 \, {\rm GeV}$, we choose the couplings to give
the largest induced splitting consistent with other experimental
constraints ($\eta_U = 0.3,\eta_D = 0.45$), $M_I = 150 \, {\rm
GeV}$ (its value before the perturbative correction in the high
scale $\rm SU(2)_C$ preserving scenario) and $M_R = (190,230) \,
{\rm GeV}$ which are the maximum values consistent with EWPD for
the $(68 \%,95 \%)$ regions. We find that the EWPD regions begin
to have tuning for a high scale degenerate mass spectrum at  $(90
\, {\rm TeV}, 8000 \, {\rm TeV})$.  Conversely choosing the
unknown $\eta_U,\eta_D \sim 0.35^2/(16 \,\pi^2)$ one finds that
the $(68 \%,95 \%)$ regions begin to have some degree of tuning
for scales of  $(170 \, {\rm TeV}, 19000 \,{\rm TeV})$. For a UV
completion that approximately preserves $\rm MFV$ and $\rm
SU(2)_C$, considering a SM and octet low energy scalar mass
spectrum allowed by EWPD is not a fine tuned scenario.

\section{Phenomenology}

We next turn to the various observational constraints. As we shall
see, the most robust constraints are those coming from the absence
of direct pair-production at LEP, which require
\be \label{LEPprod}
 M_\pm \gsim 100 \; \hbox{GeV}
 \quad \hbox{and} \quad
 M_R + M_I \gsim 200 \; \hbox{GeV} \,.
\ee
Since the octet scalar couples to both photons and gluons, these
constraints are essentially kinematic up to the highest energies
probed by LEP (more about which below).

\subsection{Fits to Electroweak Precision Data}

\begin{figure}[h]
\centerline{\scalebox{1}{\includegraphics{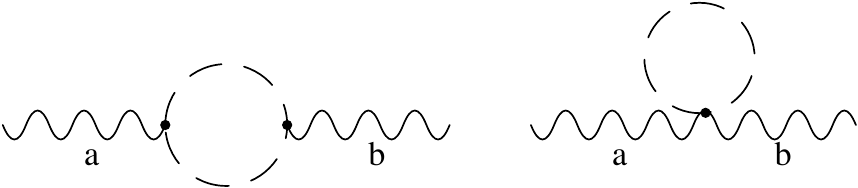}}}
\caption{Self energies calculated for the EWPD constraints on the
octets, where $(a,b) = (W^+ W^-, Z Z, \gamma \gamma, Z \gamma)$. The self energies needed to determine $\rm STUVWX$ are
given in the Appendix. } \label{figure1}
\end{figure}

A strong restriction on the properties of exotic scalars comes
from precision electroweak measurements, whose implications we now
explore in some detail. The dominant way that such scalars
influence the electroweak observables is through their
contributions to the gauge boson vacuum polarizations; the
so-called `oblique' corrections \cite{Holdom:1990tc,
Peskin:1991sw,Altarelli:1990zd}. The calculation of the oblique
corrections proceeds as usual with the vacuum polarizations being
determined directly by evaluating the diagrams given in Figure
\ref{figure1}.

When evaluating these it is important to keep in mind that the
direct production constraints, eq.~\pref{LEPprod}, can allow one
of $M_R$ or $M_I$ to be significantly lower than 100 GeV. This is
important because it precludes our using the most commonly-used
three-parameter (S, T and U) parametrization of the oblique
corrections \cite{Holdom:1990tc, Peskin:1991sw, Altarelli:1990zd},
since these are based on expanding the gauge boson vacuum energies
out to quadratic order: $\Pi_{ab}(q^2) \simeq A_{ab} + B_{ab}
q^2$, where $a$ and $b$ denote one of $Z$, $W$ or $\gamma$. Since
the electroweak precision measurements take place at $q^2 \simeq
0$ or $q^2 \simeq M_Z^2$, using the quadratic approximation for
$\Pi_{ab}(q^2)$ amounts to neglecting contributions that are of
relative order $M_Z^2/M^2$, where $M$ is the scale associated with
the new physics of interest (in our case the new-scalar masses).
This approximation becomes inadequate for $M$ below 100 GeV, and
so we must instead use the full 6-parameter description ($\rm
STUVWX$), such as in the formalism of ref. \cite{Burgess:1993mg,
Maksymyk:1993zm}. In general, the $\rm STUVWX$ formalism reduces
to the three-parameter STU case when all new particles become very
heavy.

For ease of comparison with past results we start by quoting the
results we obtain for the fit to the six parameters of the $\rm
STUVWX$ oblique formalism, regardless of how they depend on the
parameters of the Manohar-Wise model. The results are given in
Table 1, which compares the results obtained by fitting $34$
observables (listed in an appendix) to $(i)$ all six parameters
(STUVWX); $(ii)$ only three parameters (STU); or just two
parameters (ST). The number of degrees of freedom in these fits to
$(6,3,2)$ parameters is ${\it v} = (28,31,32)$, respectively. The
$\chi^2/{\it v}$ for the three fits is within one standard
deviation $\sqrt{2/{\it v}}=(0.27, 0.25, 0.25)$ of the mean of
$1$, indicating a good quality of fit. The experimental values and
theoretical predictions used are given in Table 2 in the
Appendix.

\vspace{0.5cm}
\begin{table}[h]
  \label{table1}
\begin{center}
\begin{tabular}{|c||c||c||c|} \hline
Oblique &  STUVWX Fit ($ \chi^2/{\it v} = 0.91$)  & STU Fit ($ \chi^2/{\it v} = 0.99$)
& ST Fit ($ \chi^2/{\it v} = 0.98$) \\
\hline \hline
 S &  $ 0.07 \pm 0.41 $&$ -0.02 \pm 0.08 $& $-9.9 \times 10^{-3} \pm 0.08$\\
 T &  $ -0.40 \pm 0.28 $&$ -0.02 \pm 0.08 $& $1.1 \times 10^{-2} \pm  0.07 $\\
 U &  $ 0.65 \pm 0.33 $& $ 0.06 \pm 0.10 $& - \\
 V &  $ 0.43 \pm 0.29 $& - & - \\
 W & $ 3.0 \pm 2.5 $&  - & -\\
 X &  $-0.17  \, \pm 0.15 $& - & - \\
  \hline
\end{tabular}
\caption{EWPD Fit Results in various schemes for the $34$
observables listed in the Appendix. The $\rm STU$ and $\rm ST$
fits fix the other oblique corrections to zero as a prior input.
The error listed is the square root of the diagonal element of the
determined covariance matrix. The central values of the fitted
oblique corrections  decrease as more parameters are turned off.
All three fits are consistent with past results and the PDG quoted
fit results.
}
\end{center}
\end{table}
\vspace{0.5cm}

The correlation coefficient matrix for the three fit results are
as follows,
\ba\label{cov1}
M_{STUVWX} =\left(
\begin{array}{cccccc}
 1 & 0.60& 0.38 & -0.57& 0 & -0.86 \\
 0.60 & 1 & -0.49 & -0.95 & 0& -0.13 \\
 0.38 & -0.49& 1 & 0.46 & -0.01&-0.76  \\
 -0.57 & -0.95 & 0.46 & 1 & 0 & 0.13 \\
0& 0 & -0.01 & 0 & 1 & 0 \\
  -0.86 & -0.13 & -0.76 & 0.13 & 0 & 1
\end{array}
\right),
\ea
\ba\label{cov2}
M_{STU }=\left(
\begin{array}{ccc}
 1 & 0.84& -0.20 \\
 0.84 & 1 & -0.49 \\
 -0.20 & -0.49 & 1
\end{array}
\right), \quad \quad
M_{ST} = \left(
\begin{array}{cc}
 1 & 0.87 \\
 0.87 & 1
\end{array}
\right).
\ea

\begin{figure}[h]
\centerline{\scalebox{0.65}{\includegraphics{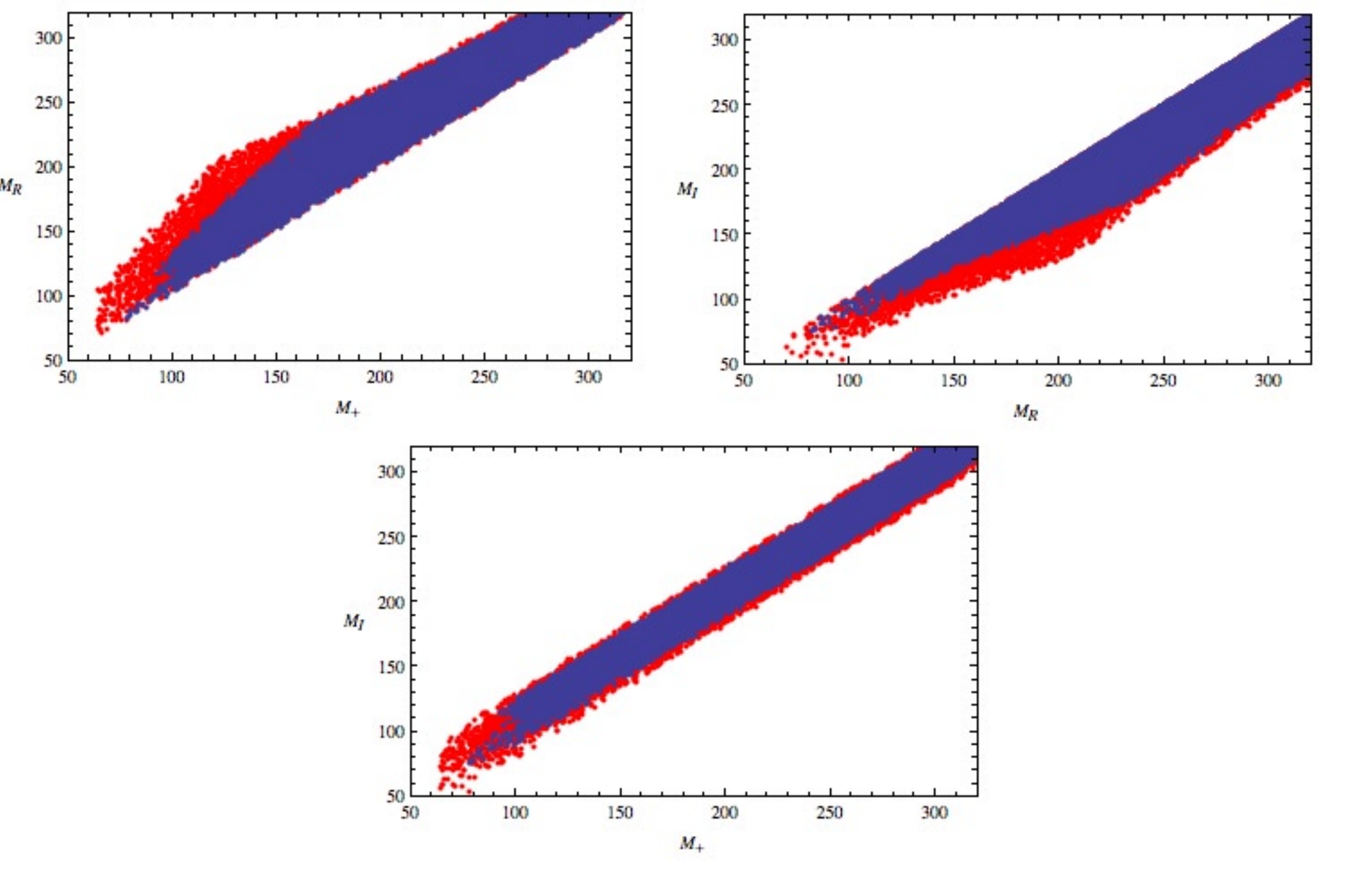}}}
\caption{Comparison of the three and six parameter fits for low
masses. (The upper two panels are not symmetric about $M_I = M_R$
and $M_R = M_+$ because we scan only through positive values for
the couplings, $\lambda_i$.) The three parameter fit is red (grey)
and the six parameter fit is blue (black). Contrary to naive
expectations the six parameter fit is more constraining on the
model despite the extra parameters; the correlations between the
extra parameters ($\rm S,X$ and $\rm U,X$ and $\rm T,V$) increases
the constraints on the model. The masses are in $\rm GeV$. EWPD
constrains the mass spectrum to be approximately $\rm SU(2)_C$
symmetric in either case where $M_{\pm} \approx M_I$. }
\label{figure2}
\end{figure}

We use the results of this fit to constrain the masses allowed in
the Manohar-Wise model by computing the vacuum polarizations as
functions of the masses of the octet and Higgs scalars. We obtain
allowed mass ranges for the scalars by demanding that the
contribution of the new physics (and the difference between the
floating Higgs mass and its fiducial value, which we take from the
SM best fits to be 96 GeV), $\Delta \, \chi^2$ which satisfies
\begin{eqnarray}\label{deltachiconstraint}
 (\C^{-1})_{i,j}  (\Delta \theta_i) \, (\Delta \theta_j)
 < 7.0385 \, (12.592)
\end{eqnarray}
for the $68 \%$ ($95 \%$) confidence regions defined by the
cumulative distribution function for the six parameter fit. Here
$\C$ is the covariance matrix constructed from the correlation
coefficient matrix given in eq.~(\ref{cov1}) or (\ref{cov2})
\begin{eqnarray}
(\C^{-1})_{i,j} = \frac{1}{2} \, \frac{\partial^2 \,
\chi^2(\theta)}{ \partial \, \theta_i \, \partial \, \theta_j}
\vert_{\theta_i = \hat{\theta_i}}
\end{eqnarray}
and $\Delta \, \theta_i = A_i - A_i^{fit}$ is the difference in
$A_i = {S,T,U,V,W,X}$ as a function of octet masses and the best
fit value, given in Table 1.

\begin{figure}[h]
\centerline{\scalebox{1}{\includegraphics{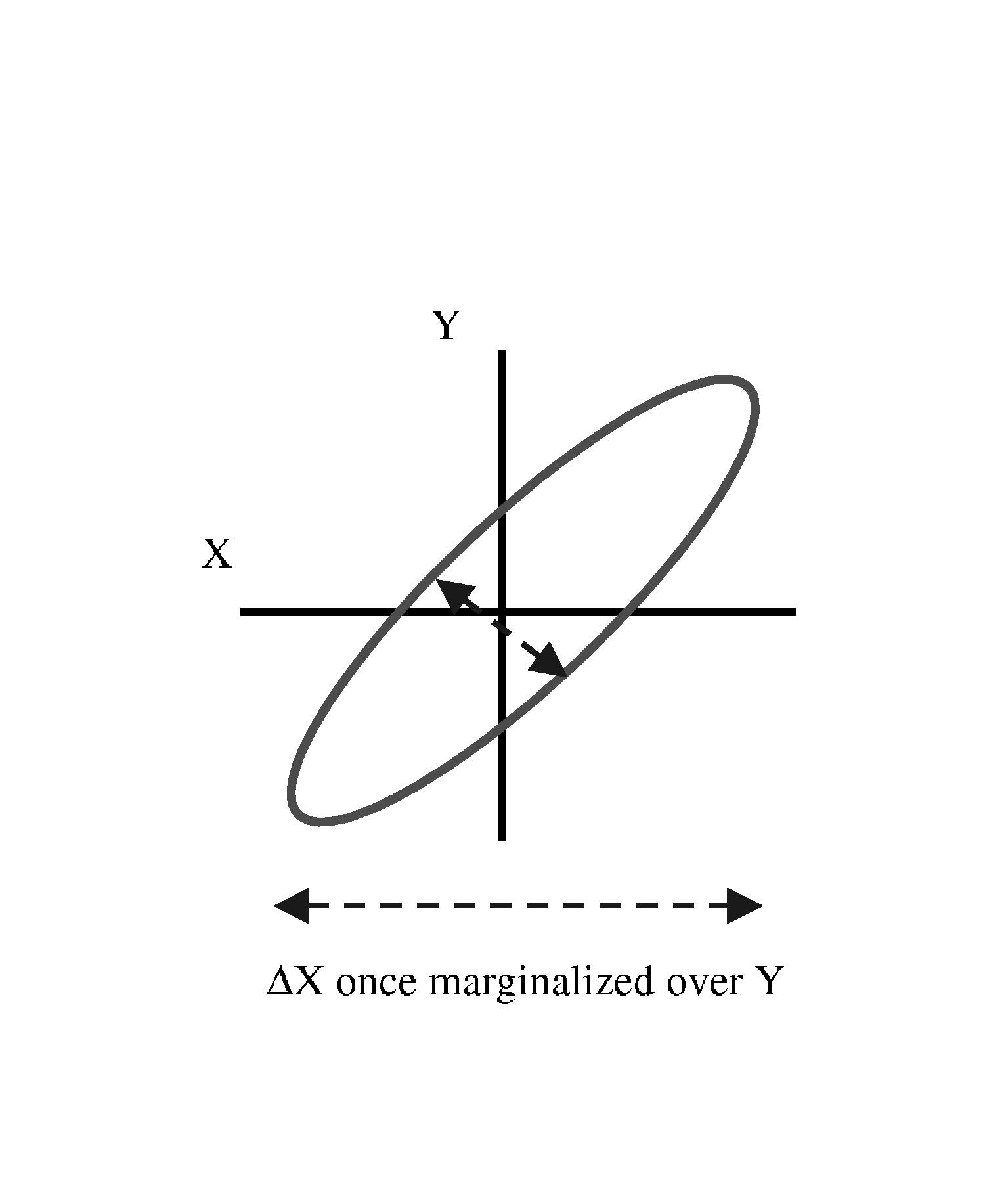}}}
\caption{A cartoon of the best-fit confidence interval for a
strongly correlated pair of variables, indicating how the best
constraints can be missed once one of the variables is
marginalized. } \label{figureellipse}
\end{figure}

An example of the best-fit regions for the allowed octet masses is
given in Figure \ref{figure2}, which compares the quality of the
constraints that are obtained using the full six-parameter
(STUVWX) parametrization, as opposed to the three-parameter (STU)
expression. The three panels plot the masses of the components of
the octet that lie within the 68\% confidence ellipsoid of the
best-fit value as the various scalar couplings, $\lambda_i$, are
varied. The two panels of this plot show how these masses are
correlated by the condition that the predictions agree with the
precision electroweak measurements, and the points in the upper
two panels all satisfy $M_I \le M_R$ and $M_+ \le M_R$ because we
choose to scan only through positive values of the couplings
$\lambda_i$.

The strongest correlation is between $M_I$ and $M_+$, for which
agreement with EWPD demands these two masses cannot be split by
more than about 50 GeV. This is as might be expected given that this
difference must vanish in the limit that the potential is
custodial invariant. The breaking of $\rm SU(2)_C$ generically
leads to bad fits because custodial-breaking quantities like the
parameter $\rho - 1 = \alpha\, T$ are measured to be very small. New sources of $\rm SU(2)_C$
breaking are extremely constrained, the deviation of the $\rho$ parameter from its SM value
is given by $\rho_0 = 1.0004^{+ 0.0008}_{-0.0004}$ \cite{PDBook}.

The comparison in Figure \ref{figure2} also shows that the
six-parameter STUVWX fit agrees with the three-parameter STU fit
when all scalars are heavy, as might be expected. It also shows
that the six-parameter fit is the more constraining one when the
octet masses are light. We understand that this happens because of
the strong correlations amongst the oblique parameters, which
implies that the best-constrained parameter direction is not
aligned along any of the STUVWX axes, as shown in Figure
\ref{figureellipse}. As a result the constraint obtained by
restricting to the axes $V = W = X = 0$ can be weaker than the
full result, significantly affecting the determined $68 \%$
confidence regions. For this reason our remaining results quote
only the results of the full six-parameter fit.

\begin{figure}[h]
\centerline{\scalebox{0.65}{\includegraphics{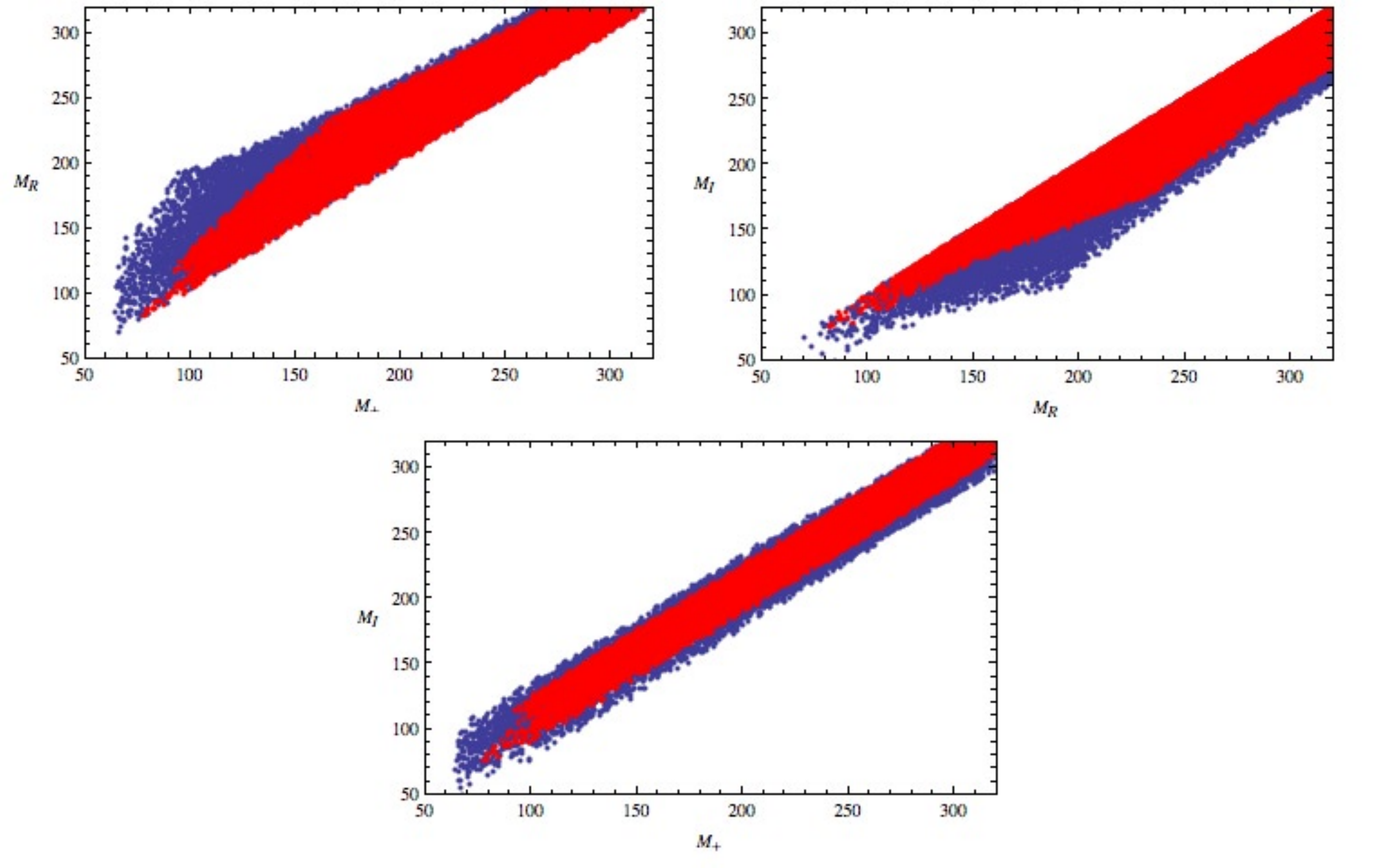}}}
\caption{Comparison of the $68 \%$ red (grey) and $95\%$  blue
(black) confidence regions when $0 < \lambda_i  <1$. The masses
are in $\rm GeV$, and $M_I, M_+ \le M_R$ because we scan only
through positive values of the couplings $\lambda_i$. For low
masses the $95\%$ confidence region is significantly expanded
compared to the $68 \%$ region, this is due to the spread of
available masses being larger for low masses, as the mass
splitting between the states scales as  $\sim v^2/m_s$. We examine the naturalness of this mass spectrum in Section 2.2 and find that
it is not simply a fine tuned solution for an
underlying new physics sector.}
\label{figure3}
\end{figure}

\subsubsection{Constraints on Octet scalars}\label{EWPDcons}

\begin{figure}[h]
\centerline{\scalebox{0.65}{\includegraphics{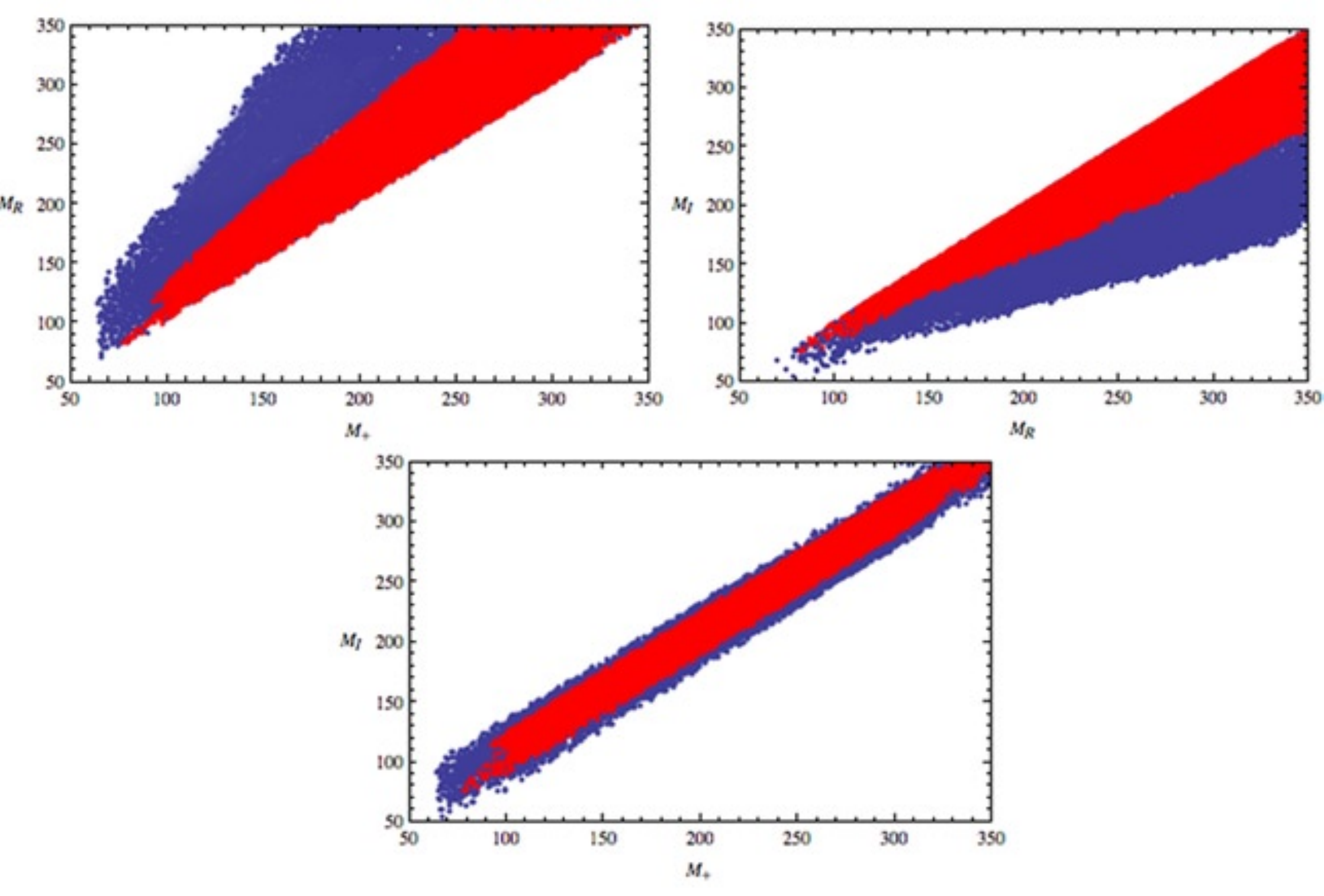}}}
\caption{Comparison of the $68 \%$ red (grey) and $95\%$ blue (black) confidence regions when $\lambda_i  <10$. Notice that the region selected for by EWPD for $M_I \approx M_+$ that is approximately $\rm SU(2)_C$ symmetric is not enlarged.}
\label{figure4}
\end{figure}

Figure \ref{figure3} displays the $68\%$ and $95\%$ confidence
regions of the model for couplings that range through the values
$0 < \lambda_i < 1$, while Figure \ref{figure4} does the same for
couplings that run through the larger range $0 < \lambda_i < 10$,
where $i=1,2,3$. As noted above, agreement with the EWPD selects
an approximately $\rm SU(2)_C$ symmetric mass spectrum, where
$\lambda_2 \approx 2 \lambda_3$ and $|M_{\pm} - M_I| < 50\, {\rm
GeV}$, but this is easily understood. Consider the case where the
octets are heavy, $v^2/M_S^2 \ll 1$, which was examined in
\cite{Manohar:2006ga}. In this mass regime it is the model that
constrains the mass spectrum to be degenerate, $M_{\pm} \approx
M_R \approx M_I$, since the mass splittings scale as $v^2/M_S$
from Eq.~(\ref{treemass}). The contribution of the octets to the S
and T parameters,\footnote{We have checked that our results in the
$\rm STUVWX$ formalism reduce to these results when $v^2/M_S^2 \ll
1$.} is then \cite{Manohar:2006ga}
\ba
 S = \frac{\lambda_2 \, v^2}{6 \, \pi \, M_S^2},
 \quad \quad \quad T = \frac{v^4}{96 \,
 \pi^2 M_S^2 s_W^2 M_W^2}(\lambda_2^2 - (2 \, \lambda_3)^2),
\ea
where $s_W \equiv \sin(\theta_W)$. Large corrections to $S$ and
$T$ are avoided if $\lambda_i$ decreases and preserves approximate
$\rm SU(2)_C$ as $M_S$ decreases, therefore allowing smaller octet
masses.

How natural are the small intra-octet splittings favoured by EWPD?
If the mass splitting is induced by the potential, while $v \gg
M_s$, for the octet masses to be allowed by EWPD that selects for
a mass degeneracy $\Delta \, M = M_I - M_{\pm}$, one would have to
require that the couplings the the octet-Higgs potential satisfy
the scaling rule
\bea\label{scal1}
 \lambda_2 - 2 \lambda_3 \ll 4 \,  \frac{\Delta \, M}{v} \, \sqrt{\lambda_1}.
\eea
As EWPD requires $\Delta \, M \sim 50 \, {\rm GeV}$ for the $95 \,
\%$ confidence region this is a mild hierarchy of couplings given
by $ \lambda_2 - 2 \lambda_3 \ll 0.8 \, \sqrt{\lambda_1}$.
Conversely for the case $m_S \gg v$, one requires that the
couplings the the octet-Higgs potential satisfy the scaling rule
\bea\label{scal2}
 \lambda_2 - 2 \lambda_3 \ll 8 \, \frac{(\Delta \, M) \, m_S}{v^2},
 \eea
which is easily satisfied for small $\lambda_i$ (which we see
below are favoured by Landau pole constraints).

The calculations presented in previous sections for the running of
custodial-breaking couplings can be used to frame a criteria as to
whether the above coupling pattern is natural. The scale
dependence of the masses is used to estimate what the $\rm
SU(2)_C$ splitting of the masses should be in the theory below the
UV scale, $\Lambda$, without tuning. One determines how high the
scale $\Lambda$ can be before the EWPD mass regions are excluded.
This quantifies the degree of fine tuning of the masses for this
scenario.\footnote{To determine the mass splitting, we technically
need to diagonalize the $S_I$ field which mixes at one loop  with
$S^R$. As the non diagonal terms in the mass matrix are one loop,
the effects of this diagonalization on the mass eigenstate $S_I'$
shifts the mass at two loop order. See the Appendix for a
determination of the mixing angle. Thus to one loop order one can
just take the one loop corrections  to $M_I$ and $M_{\pm}$ of the
last two sections, properly renormalized, to determine the mass
splitting through the counterterms.} Since the electroweak
hierarchy problem argues that the scale of new physics is likely
not too much larger than the TeV regime, we find that the favoured
mass splittings are natural, provided that the underlying theory
approximately preserves $\rm MFV$ and $\rm SU(2)_C$.

The above ranges of allowed splittings amongst scalar masses
directly constrain the three couplings $\lambda_{1,2,3}$ to be
small. But small $\lambda_i$, for $i \ge 4$, are also favoured due
to considerations of the effect of these $\lambda_i$ on the
running of the Higgs self coupling \cite{Gerbush:2007fe}. The mild
assumption that one not encounter a Landau pole while running the
Higgs self coupling up to $10 \, {\rm TeV}$, when one assumes
$\lambda_{i\ge 4} =0 $ and $m_h = 120 \, {\rm GeV}$, gives the
constraints \cite{Gerbush:2007fe}
\ba
 \lambda_1 \lesssim 1.3, \quad \quad  \quad
  \sqrt{\lambda_1^2 + \lambda_2^2} \lesssim 2.2.
\ea

However, generically $\lambda_{i\ge 4} \neq 0 $ and if the octets
and the Higgs were part of a new sector then the cut-off scale
could be lower that $10 \, {\rm TeV}$. For these reasons we only
take these constraints to inspire the $\lambda_i < 1$ limit for
the parameter space searches in Figure \ref{figure3}, but also
examine parameter space where we relax this bound to $\lambda_i <
10$ in Figure \ref{figure4}. We emphasize that direct production
bounds on the octets that rely on their fermionic decays
essentially constrain the MFV proportionality factors $\eta_i$,
while EWPD is complementary in that it  constrains the parameters
in the potential, $\lambda_i$, by constraining the mass spectrum.

\subsubsection{Implications for the inferred Higgs mass}\label{octhiggsfit}

Adding the new octet scalar to the SM also affects the best-fit
value of the Higgs mass that emerges from fits to EWPD. In
particular, we now show that the presence of the octet can remove
the preference of the data for a light Higgs, even if the new
octet scalar is also heavy.

To determine this effect we calculate the one-loop Higgs
contribution to the six oblique parameters and jointly constrain
the Higgs mass and the octet masses in the fit. For example, $S$
in this case becomes
\bea
 S = S_{oct}(M_R,M_I,M_{\pm}) +  S_{Higgs}(M_h)
 - S_{Higgs}(M_h= 96 \, {\rm GeV})
\eea
where $S_{oct (Higgs)}$ is the one-loop octet (Higgs) contribution
to the S parameter. We neglect the two-loop dependence on the
Higgs mass in the fit and this leads to an underestimate of the
allowed parameter space, as we find the $68 \% \, (95 \%)$
confidence level values of fitting the Higgs mass alone are given
by $112 \, (160) {\rm GeV}$. This gives a conservative range when
comparing to the various allowed values that are strongly
dependent on the priors used in the PDG .

\begin{figure}[h]
\centerline{\scalebox{0.65}{\includegraphics{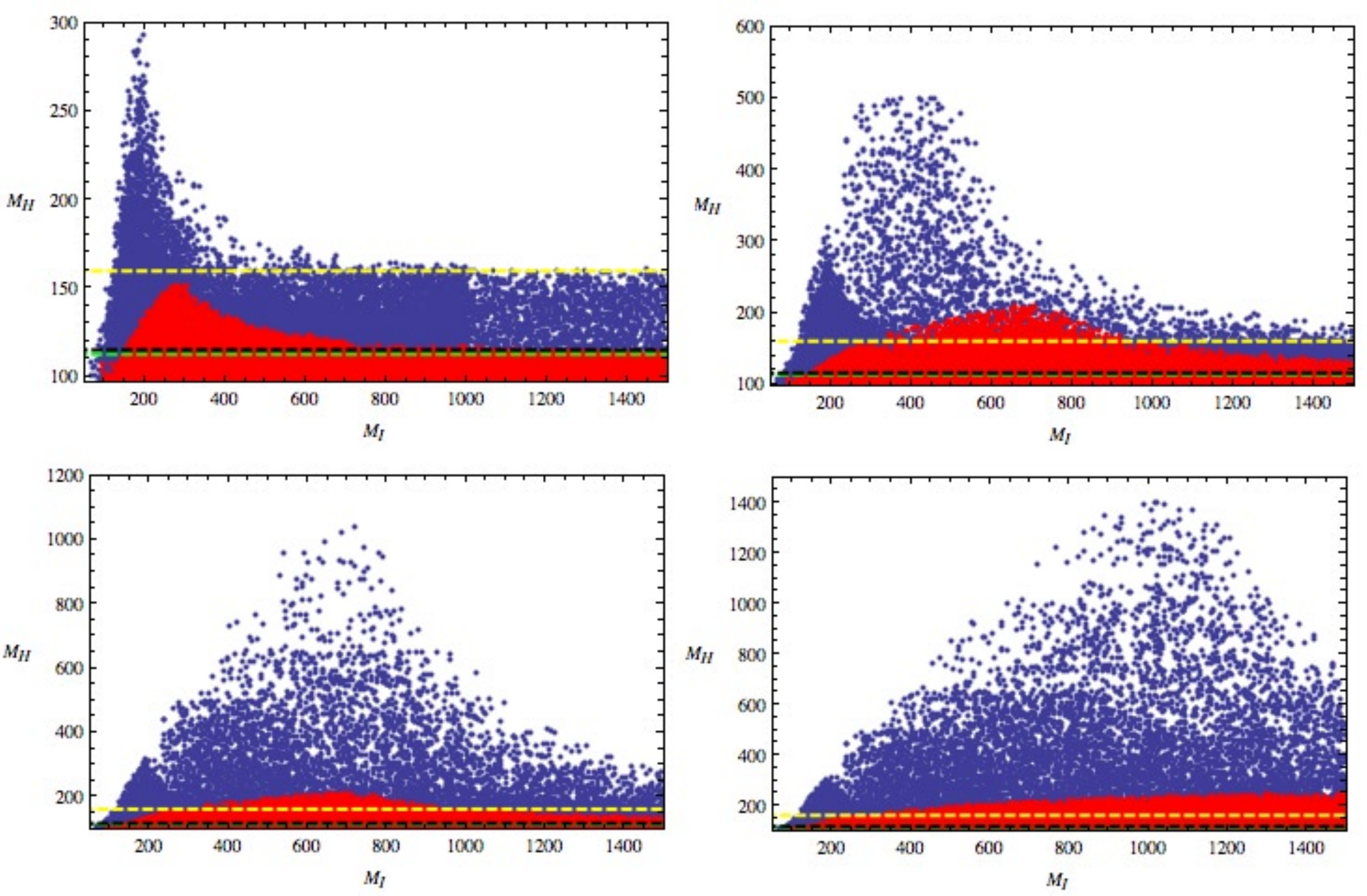}}}
\caption{The effect of octets on the fitted value of the Higgs
mass.The plots of $M_h$ versus the other octet states are
substantially the same. The green line is the $68 \%$ confidence
bound where the Higgs alone is varied at one loop. The yellow line
is the $95 \%$ confidence bound where the Higgs alone is varied at
one loop, and the black line is the direct production bound on the
Higgs mass at $95 \%$ confidence. The red (grey) region is the $68
\%$ confidence region, while the blue (black) region is the ($95
\%$) confidence region for a joint fit to the octets and the
Higgs. Notice the increase in vertical scale for the diagrams as
the upper limit of the $\lambda_i$ is increased through $1$ (upper
left), $3$ (upper right), $6$ (lower left) and $10$ (lower right).
The mechanism that is allowing the Higgs mass to increase and
still be in agreement with EWPD is the postitive $\Delta T$
contribution from the octets that is discussed in Section 3.1.2.}
\label{figure5}
\end{figure}

The effect of the octets changes the preferred Higgs mass
significantly, and two mechanisms are at work depending on the
size of the octet mass. If the octet mass $M_I$ is small, it can
allow the Higgs mass to increase by effectively replacing it in
the oblique loops, thereby giving agreement with EWPD. This is
illustrated in the upper-left plot of Figure \ref{figure5}, which
shows how a large Higgs mass correlates with small $M_I$.

The other panels of Figure \ref{figure5} reveal another mechanism
at work, however \footnote{Note that we expect a careful study of the non oblique
Higgs and octet mass dependence of $R_b$ will further constrain this parameter space with all scalars heavy but not remove it.}. In these one sees that as the upper limit on
$\lambda_i$ is increased, the upper limit on the Higgs mass
confidence regions becomes significantly relaxed. This is due to a
cancellation between the effects of the heavy octet and the Higgs
in their contributions to oblique parameters, that is made
possible by a positive $\Delta T$ contribution that the octets
give to $\chi^2$. For the three-parameter fit, the $\chi^2$ test
is of the form
\begin{eqnarray}
 (\C^{-1})_{i,j}  (\Delta \theta_i) \, (\Delta \theta_j)
 =  596 \, (\Delta S)^2 - 1159 \, (\Delta S) \,
 (\Delta T) + 751 \,  (\Delta T)^2
 \label{STchisq}
\end{eqnarray}
where we neglect contributions that are not logarithmically
sensitive to the Higgs mass at one loop, since this is all that is
relevant to the argument. For the three-parameter fit, the $68 \%$
confidence region is defined by $(\C^{-1})_{i,j}  (\Delta
\theta_i) \, (\Delta \theta_j) < 3.536$ and is easily satisfied
for light Higgs masses. As the Higgs mass grows, its contribution
to $(\Delta S)$ and $(\Delta T)$ becomes dominated by the
logarithmic dependence
\be
 (\Delta S) \simeq \frac{\alpha}{12 \, \pi}
 \log \left(\frac{M_H^2}{\hat{M}_H^2} \right)
 \quad \hbox{and} \quad
 (\Delta T) \simeq - \frac{3 \, \alpha}{16 \, \pi}
 \log \left(\frac{M_H^2}{\hat{M}_H^2} \right),
\ee
where $\hat{M}_H$ is the reference value of the Higgs mass, which
for our fit is $96 \, {\rm GeV}$. The crucial point is that
$(\Delta T)$ is negative for $M_H > \hat{M}_H$ and for the SM this
quickly excludes large Higgs masses because of the sign flip in
the $(\Delta S) \, (\Delta T) $ term in $\chi^2$.

Including the contribution of the octets in the large mass regime
($v^2/M_S^2\ll1$) modifies these expressions to
\bea
 (\Delta S) &\simeq& \frac{\alpha}{12 \, \pi}
 \log \left(\frac{M_H^2}{\hat{M}_H^2} \right)
 +  \frac{\lambda_2 \, v^2}{6 \, \pi \, M_S^2},\nn \\
 (\Delta T) &\simeq& - \frac{3 \, \alpha}{16 \, \pi}
 \log \left(\frac{M_H^2}{\hat{M}_H^2} \right) +
 \frac{v^4}{96 \, \pi^2 M_S^2 s_W^2 M_W^2}(\lambda_2^2
 - (2 \, \lambda_3)^2),
\label{STocthiggs}
\eea
where the factor $\lambda_2^2 - (2 \, \lambda_3)^2$ comes from a
factor of $(M_R^2 - M_{\pm}^2) (M_I^2 - M_{\pm}^2)$ in the octet
contribution, and is a measure of the total mass splitting in the
doublet. For $\lambda_i >0$, we know $M_R^2 >  M_{\pm}^2$ and so
the octets give a positive contribution to $(\Delta T)$ so long as
$M_I^2 > M_{\pm}^2$. The octets (or any other doublet with gauge
couplings and small mass splittings) then allow $(\Delta T)$ in
Eqn.~\ref{STocthiggs} to be positive, and so allow a large degree
of cancellation between the $(\Delta S)^2, (\Delta T)^2$ and
$(\Delta S)(\Delta T)$ terms in Eqn.~\ref{STchisq}. The size of
the positive $(\Delta T)$ contribution scales with the upper limit
on $\lambda_i$, explaining the significant relaxation of the Higgs
mass bound in Figure \ref{figure5}. We find that the Higgs and the
octet scalars could both have masses $ \sim 1 \, {\rm TeV}$ and
still lie within the $95 \%$ contour mass region allowed by EWPD.
We also note that we restrict our searches to positive $\lambda_i$
(which must be so for at least some of the couplings to ensure the
absence of runaway directions in the potential), however clearly
negative $\lambda_2$ could also act to relax the EWPD bound on the
Higgs mass by giving a negative contribution to $(\Delta S)$.

We emphasize the generic nature of the mechanism, wherein the
contributions of TeV scale new physics can mask the contributions
of a heavy Higgs to electroweak precision observables. It applies
in particular when EW symmetry breaking leads to a mass splitting
of an extra $\rm SU(2)$ doublet, since the extra doublet can give
a positive contribution to $(\Delta T)$ proportional to the mass
splittings of the doublet components. This has been recognized as
a simple way to raise the EWPD bound on the Higgs mass by
satisfying the positive $(\Delta T)$ criteria of
\cite{Peskin:2001rw}. Expressed as an effect on the $\rho$
parameter, it also has a long history going back to observations
by Veltman \cite{Einhorn:1981cy}, being rediscovered for
two-Higgs-doublet models in \cite{Chankowski:2000an}, and used for
the construction of the Inert Two Higgs doublet (IDM) model
\cite{Barbieri:2006dq}.\footnote{For a similar construction see
\cite{Deshpande:1977rw}} In this latter model, the Higgs mass is
raised, addressing the "LEP paradox", and the naturalness of the
SM Higgs sector is also improved by raising the cutoff scale of
the modified SM. In the IDM model a parity symmetry is imposed to
avoid FCNC's.

We note that the example of the general scalar sector consistent
with flavour constraints, the Manohar-Wise model examined in this
paper, naturally has a number of the benefits of models like the
IDM while avoiding the imposition of a parity symmetry. Allowing
the second doublet to couple to quarks improves its potential for
detection, without introducing large FCNCs due to MFV.  It is
interesting that the effect of raising the Higgs mass has emerged
naturally from the most general MFV scalar sector and was not a
model building motivation of the MW model. Variants of the MW
model, can address the naturalness of the scalar sector through
raising the cut off scale and further the colour charge of the
octet provided some rational for the second doublet not obtaining
a $\it vev$, through the avoidance of the spontaneous breaking of
colour. Also, for the entire parameter range, octets skew the
distribution of the allowed Higgs masses so that the direct
production bound on the Higgs mass and the EWPD fit of the Higgs
mass can be in better agreement.

\subsubsection{Implications for the tension between leptonic and
hadronic asymmetries}\label{pulls}

Although the SM produces a good quality global fit to EWPD, there
exists a mild tension in the data between the leptonic and
hadronic asymmetries. In particular $A_{FB}^b$ deviates from the
SM predicition by $2.5 \sigma$ and favours a heavy Higgs $\sim 400
{\rm GeV}$, while $A_e$ differs from the SM by $\sim 2 \sigma$ and
favours a Higgs mass far below the direct production bound. Here
we address the question of whether the oblique contributions of
octet scalars can change this tension.

\begin{figure}[h]
\centerline{\scalebox{0.55}{\includegraphics{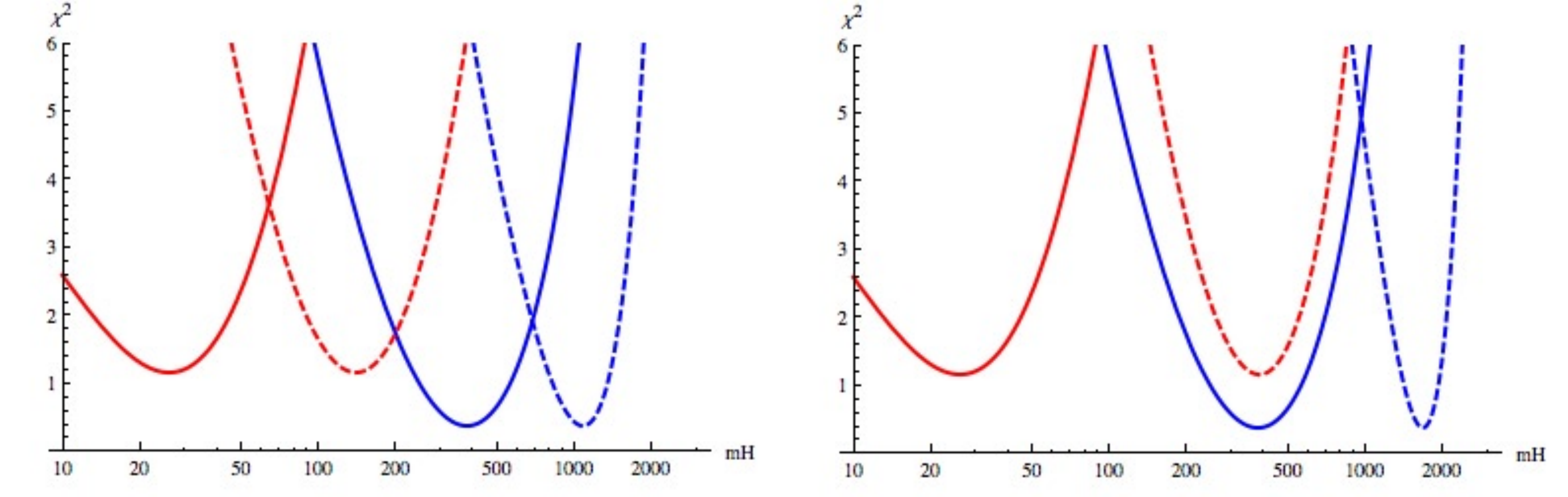}}}
\caption{The $\chi^2$ of the leptonic asymmetries (red) and
hadronic asymmetries (blue) as a function of Higgs mass in ${\rm
GeV}$. The solid curves show the contribution of the Higgs alone
and the dashed curves are for the Higgs and the octets. The figure
on the left is for octet masses $(M_\pm, M_R, M_I)=(300,400,330)
{\rm GeV}$ and on the right is for $(M_\pm, M_R,
M_I)=(900,1000,940) {\rm GeV}$.} \label{figpulls}
\end{figure}

To this end we calculate $\chi^2$ for the hadronic asymmetries
$A_{FB}^b , \,A_{FB}^c, \,A_b, \,A_c$, and for the leptonic
asymmetries using $A_\tau$ and the $A_e$ values given in Table
~\ref{data}. The results are shown in Figure \ref{figpulls}, where
the solid curves plot $\chi^2$ with the SM Higgs alone and the
dashed curves include the octets for a particular mass spectrum
allowed by EWPD. The two panels compare results for relatively
light and relatively heavy octet scalars.

The figure shows that the preferred value of the Higgs mass is
strongly dependent on the mass splitting of the octets. As
discussed in Section \ref{octhiggsfit}, the octets, unlike the
Higgs, give a positive contribution to $\Delta T$, which depends
on the mass splitting in the doublet. This increases the allowed
value of the Higgs mass. The octets can change the pull of $A_e$,
for example, to favour large Higgs masses, however they also do
the same to $A_{FB}^b$. As can be seen from Figure \ref{figpulls},
although the leptonic and hadronic asymmetries can now both prefer
a Higgs masses above the direct production bound of $114.4 \, {\rm
GeV}$, they are not brought in to closer agreement in their
predictions for the value of $M_H$.

We see from this that the octet oblique contributions do not in
themselves remove the tension between the leptonic and hadronic
asymmetries. However, because the octets are coloured it is
possible that their non-oblique corrections to $A_{FB}^b$ might be able to
bring together the leptonic and hadonic observables. We leave this
observation to a more complete calculation, which lies beyond the
scope of this paper.

\subsection{Direct-production constraints from LEP}

\begin{figure}[h]
\centerline{\scalebox{0.75}{\includegraphics{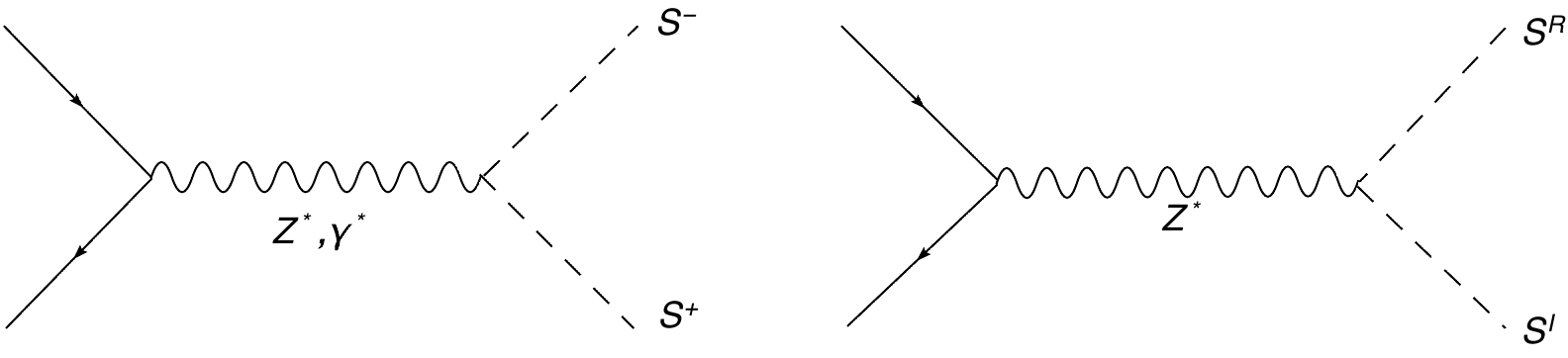}}}
\caption{
The tree level production mechanism for
$S^+ + S^-$ and $S^0_R + S^0_I$ at LEPII. }
\label{lepd}
\end{figure}

The octets would have been directly produced at LEP2 if they were
light enough through the processes in Figure \ref{lepd}.

\begin{figure}[hb]
\centerline{\scalebox{0.6}{\includegraphics{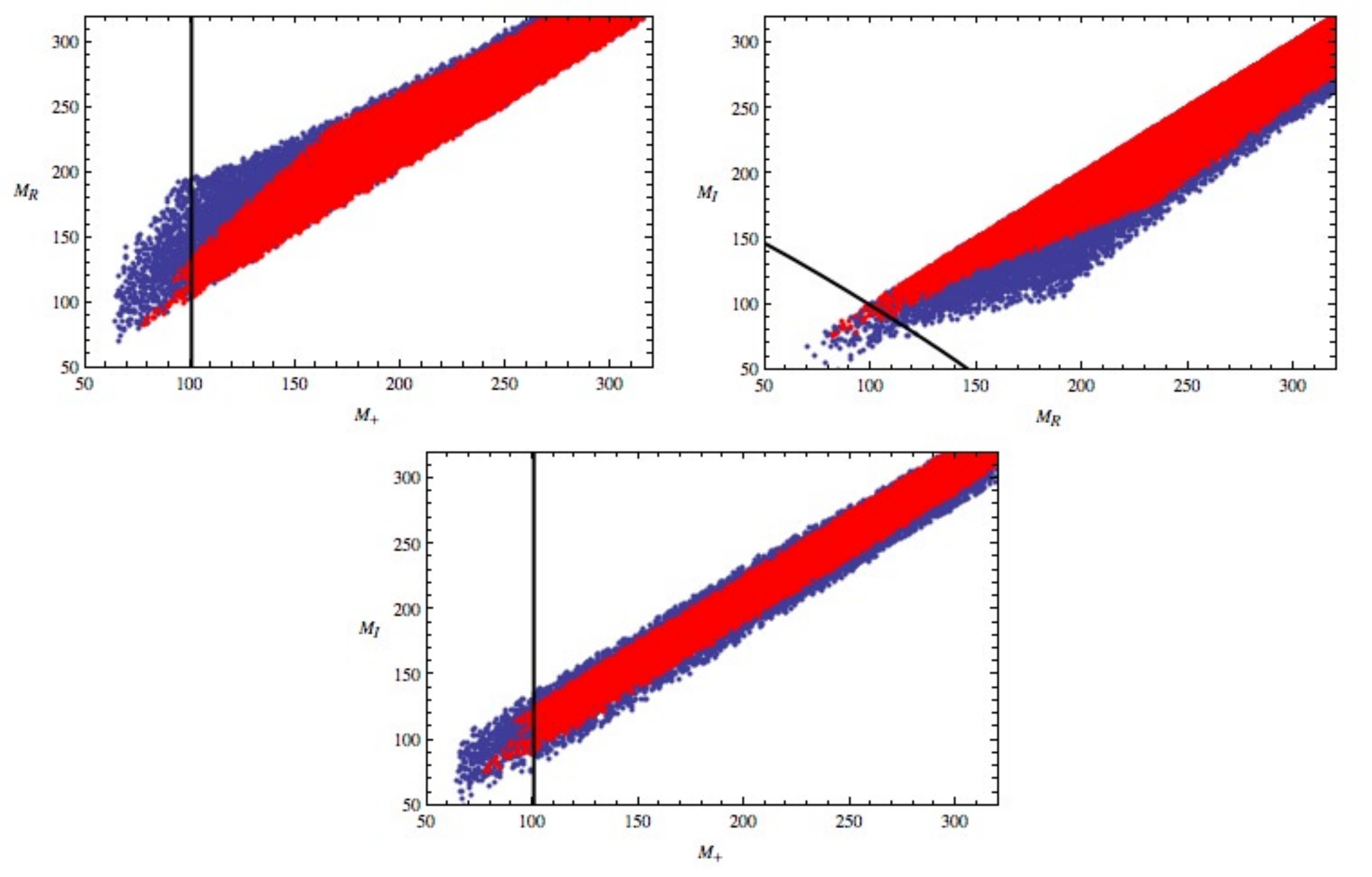}}}
\caption{ Comparison of the $68 \%$ (red or light) and $95\%$
(blue or dark) confidence regions when $\lambda_i  <1$. The LEP2
production bound for ten events is the black line.}
\label{lepoctets}
\end{figure}

The production cross sections are given by
\begin{eqnarray}
 \sigma_{S^+ S^-} &= &\frac{d_A}{4} \left(\frac{4\pi \alpha^2}{3 s} \right)\lambda^{3/2}\left(1, \frac{M_+^2}{s},\frac{M_+^2}{s} \right) \\
&\times & \left\{ 1 + 2 v_+ v_e {\rm Re}\left[ \left(1-\frac{M_Z^2}{s}+\frac{i M_Z \Gamma_Z }{s}\right)^{-1} \right] +v_+^2 (v_e^2+a_e^2)\bigg\vert 1-\frac{M_Z^2}{s}+\frac{i M_Z \Gamma_Z }{s} \bigg\vert^{-2} \right\},  \nonumber \\
&& \nonumber \\
\sigma_{S^0_R S^0_I} &= &\frac{d_A}{4} \left(\frac{4\pi \alpha^2}{3 s} \right)\lambda^{3/2}\left(1, \frac{M_R^2}{s},\frac{M_I^2}{s} \right) v_0^2 (v_e^2+a_e^2)\bigg\vert 1-\frac{M_Z^2}{s}+\frac{i M_Z \Gamma_Z }{s} \bigg\vert^{-2},
\end{eqnarray}
where we have defined $d_A= 8$, $a_e= - (4 s_W c_W)^{-1}$
\begin{eqnarray}
\lambda(x_,y,z)&=&x^2+y^2+z^2-2 x y - 2 x z- 2 y z,  \\
v_+&=& \frac{s_W^2-c_W^2}{2 s_W c_W}, \quad \quad  v_0= \frac{1}{2 s_W c_W},  \quad \quad v_e= \frac{-1 + 4 s_W^2}{4 s_W c_W}
\end{eqnarray}
The highest COM energy at which LEP2 operated was $\sqrt{s} = 209$
GeV, where approximately $\displaystyle {\int {\cal L}} dt \sim
0.1$ fb$^{-1}$ of integrated luminosity was collected. We give a
rough estimate of the sensitivity of LEP2 to light octets by
requiring less than 10 total events for a given set of masses,
$\sigma \times \displaystyle {\int {\cal L}} dt<10$. Note that
these limits are essentially kinematic limits for production, and
more accurate exclusions in the mass parameter space are possible,
but these will be dependent on the detailed decays of the octets
and SM backgrounds and be weaker constraints. The LEP2 production
bounds are shown in Figure \ref{lepoctets}.

\subsection{Tevatron constraints}

\subsubsection{Dijet constraints on the production cross section.}

Heavy octet production via gluon fusion has been examined in some
detail in the literature see
\cite{Manohar:2006ga,Gresham:2007ri,Idilbi:2009cc}. We use the
results of \cite{Manohar:2006ga, Gresham:2007ri,Idilbi:2009cc} to
determine the production cross sections for light octets  and
consider the relevant bounds on the model in this region from the
Tevatron. The single production cross section we use,
\cite{Gresham:2007ri}, neglects for simplicity the scalar mass
splitting and assumes that $\eta_U$, $\lambda_4$ and $\lambda_5$
are real. However, note that this is partially justified for
light masses as EWPD selects for an approximately degenerate mass
spectrum with an approximate $\rm SU(2)_C$ symmetry in the
underlying potential, giving $\lambda_4 =\lambda_5^\star$ and one need only assume one
of the couplings are real.
\footnote{Note that setting $\lambda_4$ and $\lambda_5$ to
real values removes the scalar loop contributions to the single
production of $S_I$, which can become large as the values of
$\lambda_{4,5}$ increases.} For the sake of simplicity we will
also neglect the effects of mixing of the $S_I$, $S_R$ states that
can occur if the effective yukawa couplings of the octet carries a
phase as discussed in the Appendix. The pair production cross
section for the charged scalars is twice that for the real scalars
\cite{Manohar:2006ga} and so is not shown.

The tree level pair production dominates the loop suppressed
single production in the low mass region for small
$\lambda_{4,5}$. However as $\lambda_{4,5}$ increase the single
production contribution takes over, which occurs at $\lambda_{4,5}
\sim 2$ for the neutral scalar, $S_R$, with a mass of $200$ GeV.

\begin{figure}[hbt]
\centerline{\scalebox{0.6}{\includegraphics{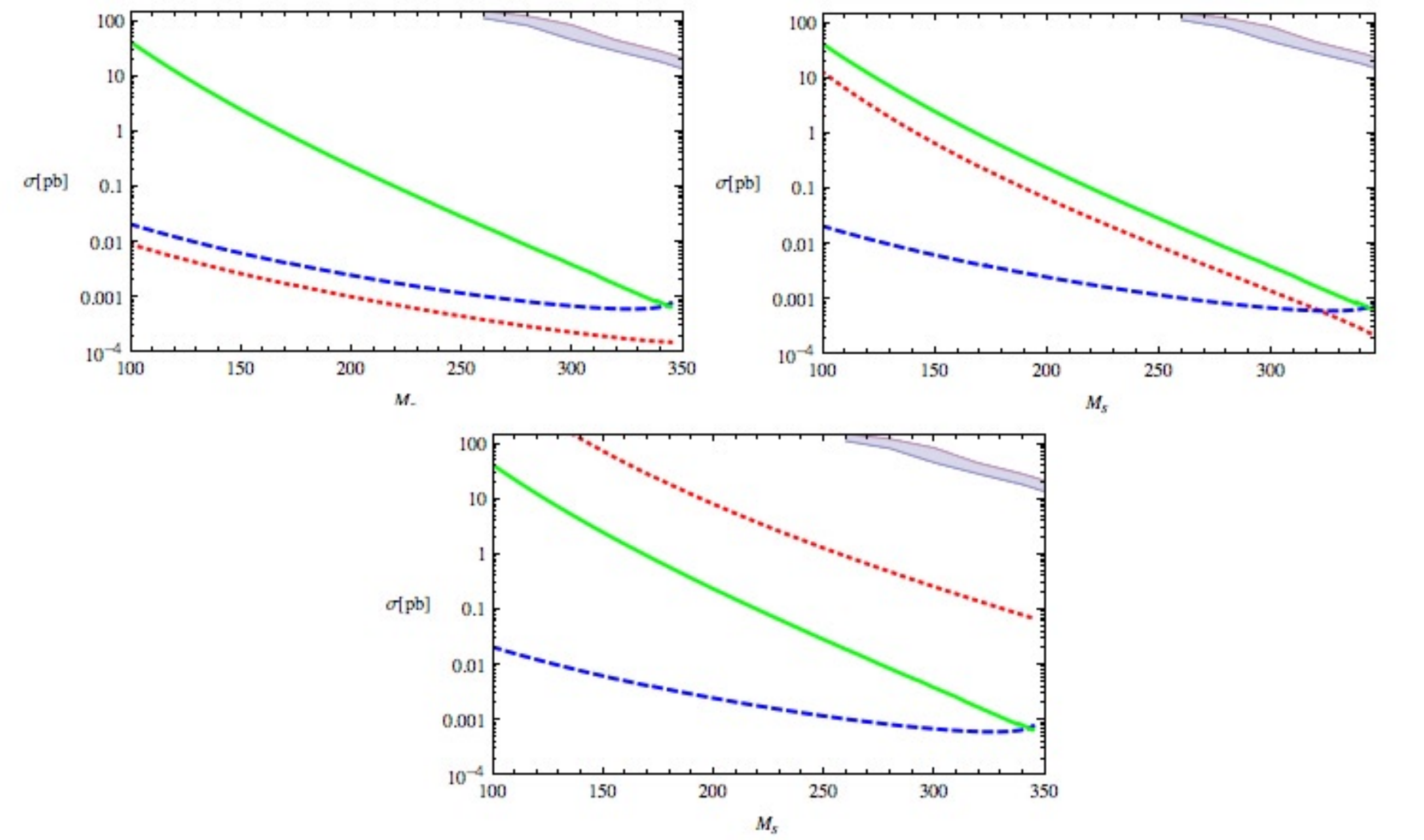}}}
\caption{Shown is the production cross section of $\sigma(g \, g
\to S_R)$ red short dashed line, $\sigma(g \, g \to S_I)$ blue
long dashed line, and the $\sigma(g \, g \to S_R \, S_R)$ given by
the solid green line. The results are for Tevatron with $\sqrt{s}
= 1.96 \, {\rm TeV}$, $\alpha_s(M_Z)=0.1217$, $m_t=173.1 \, $GeV,
$M_Z=91.1876 \, $GeV and the NLO CTEQ5 pdfs. The values of
$(\lambda_4,\lambda_5)$ chosen are $(0,0)$ upper left, $(1,1)$
upper right and $(10,10)$ for the bottom graph. In all three
graphs we have set $\eta_U = 0.2$. The dependence on $\eta_U$ is
weak and as $\eta_U$ decreases the production cross sections
decrease. Also shown is a $95 \% $ confidence limit band (the
shaded region) derived from \cite{Aaltonen:2008dn}  that places an
upper bound on new physics that decays to dijets. The region is
defined by the upper limit on $\sigma(X) \B(X \to jj)*A(|y| <1)$
where the difference between the $W'$ and RS graviton $G^\star$
$95 \% $ confidence upper bounds are taken and the acceptance
fraction requires the leading jets to have rapidity magnitude $|y|
<1$. The exclusion region depends weakly on the shape of the
resonance, so a dedicated study is required to exactly bound the
octet decay to dijets, however, the octet signal is orders of
magnitude below the exclusion regions obtained from Tevatron
before branching and acceptance ratios further reduce the signal.
A resummation of large threshold logarithms for single $S$
production was performed in \cite{Idilbi:2009cc}. The K factors
for single $S$ production was found to be $\sim 2$ for $500 \, \rm
GeV$ a octet mass  and this K factor falls as the mass decreases.
This indicates that threshold enhancements will not raise the
cross section enough to exclude octets for the entire low mass
region. } \label{figure10}
\end{figure}

A direct search strategy to find octets is to look for narrow
resonance structures above the QCD background for states that
decay into dijets. CDF has recently performed such a search
\cite{Aaltonen:2008dn} with $1.13 \, fb^{-1}$ of data that could
discover octet bound states \cite{Kim:2008bx} or single $S_i$ that
decay to dijets above the QCD background. The cross sections for
the production of these states at the Tevatron, leading to dijet
resonance structures, are orders of magnitude below the QCD
background in the regions of parameter space we consider, this is
shown in Fig. \ref{figure10}

The low mass region is not directly ruled out, although a
dedicated study to refine the lower mass bound is warranted due to
the shape dependence of the exclusion bound.\footnote{Other
possible indirect search strategies for the effects of octet
scalars include determining the effect of the octets on the
$A^t_{fb}$. In a similar manner to axigluons
\cite{Antunano:2007da}, these new exotic coloured states could
contribute to $A_{fb}^t$ as they are coloured, couple strongly to
tops, and are not a vectorlike state. Interestingly, $A_{fb}^t$
has recently been measured \cite{Aaltonen:2008hc,CDF9724} to be
$A_{FB}^t = 0.19 \pm 0.065 (stat) \pm 0.024 (syst)$ which is a
deviation larger than 2 sigma from its SM value
\cite{Antunano:2007da} of $A_{FB}^t = 0.05 \pm 0.015$.}

\subsubsection{Gauge boson decays and Lepton Signatures}\label{threebody}

The decays of the octets involving gauge bosons
\bea\label{leptond}
S_{R,I} &\to& W^{\pm} \, S^{\mp}, \quad  S_{R,I} \to Z \, S_{I,R}\nn \\
S^{\pm}&\to& W^{\pm} \, S_{R,I}, \quad  S^{\pm} \to Z \, S^{\pm}.
\eea
were studied in some detail in \cite{Manohar:2006ga, Gerbush:2007fe}. These decays are of phenomenological interest as the gauge bosons can be a source of leptons to trigger on at LHC and Tevatron. The EWPD constraints $|M_{\pm} - M_I|<50\, {\rm GeV}$ and for most of the allowed parameter space $|M_{i} - M_j|<M_W,M_Z$,
as the mass splitting of the doublets scale as $v^2/M_s$ for large masses. This causes the decays to proceed through an offshell gauge boson for most of the allowed parameter space. In this case an effective local operator can be used to approximate the decays. 

For example consider $S_R \to S^{-} \, \ell^+ \, \nu$ through an off shell $W$. The effective Lagrangian at leading order is given by the product of scalar octet and left handed lepton currents
\bea
 \mathcal{L_{\rm eff}} = \frac{-i \, g_1^2}{\sqrt{2} \, M_W^2} \, (S_R \, \partial_\mu S_{+}) \, (\bar{ \nu}_L \, \gamma^{\mu} \,\ell_L ).
\eea
Exact formula for three body decays such as this exist in the literature \cite{Pospelov:2008rn}. For the masses allowed by EWPD\footnote{This assumes that the initial state that is eventually triggered on is not highly boosted. This is generally the case due to the kinematic reach of the Tevatron and LHC.} generally the energy release is $\Delta = M_R - M_\pm < M_{R},M_{-}, M_W$. 
The resulting decay width at leading order in $\Delta/M_R$ is 
\bea\label{threebodyform}
\Gamma_{\ell} = \frac{\alpha^2 \, \Delta^5}{60 \, \pi \, s_W^4 \, M_W^4}.
\eea

When $M_R > 2 \, m_t$ the decays to leptons through an offshell $W,Z$ are suppressed decay channels. The dominant decay widths are to $t \, \bar{b}$, $t \, \bar{t}$ unless $\eta_U \ll \eta_D$. The ratio of $\Gamma_{\ell}$ to this decay, in the limit $M_R \gg 2 \, m_t$, is given by 

\bea
\frac{\Gamma_{\ell}}{\Gamma_{S^0_R \to t \, \bar{t}}} \simeq \frac{0.005 {\rm GeV}}{M_R \, |\eta_U|^2} \, \left(\frac{\Delta}{50 \, {\rm GeV}} \right)^5
\eea
for $\alpha = 1/128$, $s_W = 0.48$ and $m_t = 173 \, {\rm GeV}$.

When $M_R < 2 \, m_t$  the offshell $W,Z$ will be dominant decay channels for light masses for much of the parameter space. Taking $m_b = 4.23 \, {\rm GeV}$, and the other factors as before, the ratio of the offshell decay to the $S^0_R \to b \, \bar{b}$ decay is given by
\bea
\frac{\Gamma_{\ell}}{\Gamma_{S^0_R \to b \, \bar{b}}} &\simeq& \frac{4 \alpha^2}{15\, s_W^4 |\eta_D|^2} \, \left(\frac{\Delta^5 \, v^2}{m_W^4 \, m_b^2 \, M_R} \right), \nn \\
&\simeq& \frac{8 \, {\rm GeV}}{M_R \, |\eta_D|^2} \, \left(\frac{\Delta}{50 \, {\rm GeV}} \right)^5. 
\eea

If the dominant fermionic decays are to charm quarks due to a mild hierarchy of $\eta_U > (m_b/m_c) \, \eta_D$,  then taking $m_c = 1.3 \,  {\rm GeV}$ gives the branching ratio 
\bea
\frac{\Gamma_{\ell}}{\Gamma_{S^0_R \to c \, \bar{c}}} \simeq  \frac{82 \, {\rm GeV}}{M_R \,|\eta_U|^2 } \, \left(\frac{\Delta}{50 \, {\rm GeV}} \right)^5. 
\eea

Thus when quark decays are suppressed through $M_R < 2 \, m_t$ the dominant decay mode will be through an offshell $W,Z$  for much of the parameter space of $\eta_U , \eta_D$ allowed by other constraints, notably the constraints due to $R_b$. This sets a lower bound on the decay width of the heavier octet species given parametrically by Eqn. \ref{threebodyform}. This sets an upper bound on the lifetime of these components of the octet doublet of $4.5/\Delta^5 \, \,  ps$ which yields a
upper bound on the decay length of the form $10^{-3}/\Delta^5 \, \,  m$.\footnote{Here we have converted units assuming that   $\Delta$ is given in $\rm GeV$ as a pure number, ie for $\Delta = 50 \,  \rm GeV$ we have a upper bound on the lifetime of $1.2 \times 10^{-2} \, \, as$.} Thus the heavier octet species will decay promptly inside the detector and not leave a long lived charged track signature.

As dominant decay modes of the heavy components of the octet doublet (when  $M_i < 2 \, m_t$) can be three body decays, the final state signature would be excess monojet or dijet (depending on the boost of the final state octet) events in association with a lepton and missing energy, or enhancements of dilepton signatures with a monojet or dijet. Dedicated studies of these signatures are warranted. The lifetime of the lightest component of the octet doublet is dictated by its decay to fermion pairs.

\subsubsection{Constraints from $t \, \bar{t}$  decays.}

\begin{figure}[hbt]
\centerline{\scalebox{0.6}{\includegraphics{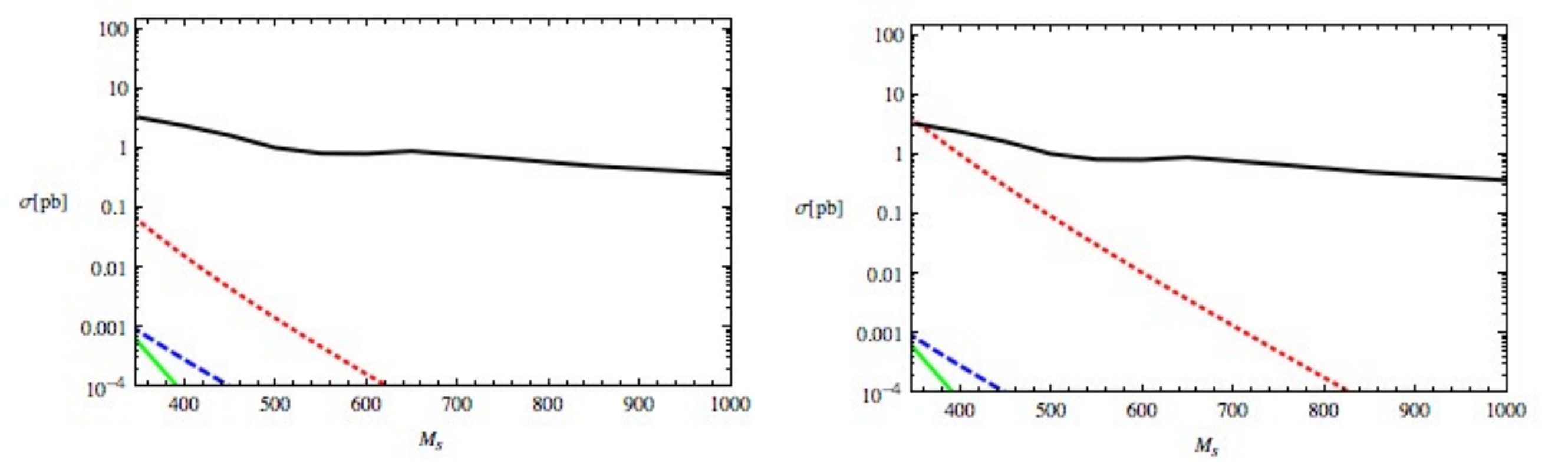}}}
\caption{Shown is the production cross section of $\sigma(g \, g
\to S_R)$ red short dashed line, $\sigma(g \, g \to S_I)$ blue
long dashed line, and the $\sigma(g \, g \to S_R \, S_R)$ given by
the solid green line. The results are for Tevatron with $\sqrt{s}
= 1.96 \, {\rm TeV}$, $\alpha_s(M_Z)=0.1217$, $m_t=173.1 \, $GeV,
$M_Z=91.1876 \, $GeV and the NLO CTEQ5 pdfs are used. The D0 $95
\%$ confidence limit on $\sigma(X) \Gamma (X \to t \, \bar{t})$ is
the upper solid black line \cite{Abazov:2008ny}. The values of
$(\lambda_4,\lambda_5)$ are $(10,10)$ for the left hand figure and
$(75,75)$ for the right hand figure. $\eta_U = 0.2$ for both
figures. For perturbative $\lambda_i \lesssim 10$, current
Tevatron production bounds on resonances in $t \, \bar{t}$ do not
rule out octets  of mass $350 - 1000 \,  {\rm GeV}$.}
\label{figure11}
\end{figure}

For neutral octet masses above $2 \, m_t$, decays into top quark
pairs can be dominant. These were previously considered in
\cite{Gresham:2007ri}. The observed limits on excess $\sigma_X \,
\cdot \, B(X \to t \, \bar{t})$ at Tevatron with $0.9 \, fb^{-1}$
of data \cite{Abazov:2008ny} do not rule out octets in the
intermediate mass region $350 - 1000 \,  {\rm GeV}$.  The
production cross section for single $gg \to S_R$ production can
become large enough for the bound on $t \, \bar{t}$ to be
relevant, however this requires $\lambda_4 \sim \lambda_5 \sim 75$
which is well into a nonperturbative region of the potential
making any conclusion suspect. We illustrate these limits in Fig.
\ref{figure11}

\subsubsection{Constraints from $\bar{b} \,  b \,
\bar{b} \,  b$ decays.}

The dominant decays for light masses will be to quarks $S_{+} \to
t \bar{b}$, $S_{R,I} \to b \bar{b}$ below the $t \bar{t}$
threshold for $\eta_{U,D} \sim \mathcal{O}(1)$. In this regime
\cite{Gerbush:2007fe} places a lower bound on the scalar mass of
approximately $200$ GeV from the CDF search for a scalar particle
decaying dominantly to $b \bar{b}$ when produced in association
with $b$ quarks \cite{CDF8954} This bound is avoided for almost
all of the available parameter space for light octet masses.
$S_{I,R}$ can decay preferentially to charms, which corresponds to
a mild hierarchy of couplings
\bea
\frac{|\eta_D|^2}{|\eta_U|^2}<\frac{m_c^2}{m_b^2} \sim
\frac{1}{10}
\eea
when neglecting $\mathcal{O}(m^2_{b,c}/M^2_S)$ terms. Neutral
scalar masses below $200$GeV are allowed for $\eta_D \lesssim
0.1$, given an upper limit of $\eta_U \sim 0.3$ from
\cite{Gresham:2007ri} for masses in this range. The three body
decays discussed in Section \ref{threebody} are actually dominant
over quark decays for much of the parameter space allowed by EWPD
for light octet masses, invalidating the assumptions of
\cite{Gerbush:2007fe} for most of the remaining parameter space.

\subsubsection{Constraints from $\gamma \, \gamma$ decays.}

A promising signature for octets at hadron colliders is the
annihilation of a pair of charged octets to photons, $ gg \to S^+
\, S^- \to \gamma \, \gamma$. We can use the recent results of DO
\cite{Abazov:2009kq,D05858} that utilize $4.2 \, fb^{-1}$ of date
to place $95 \%$ confidence upper limits on $\sigma(h) \times BR(h
\to \gamma \, \gamma)$ compared to the SM Higgs signal to directly
constrain octet annihilation into $\gamma \, \gamma$. We must
consider annihilation decays of octet bound states, octetonia,
studied in \cite{Kim:2008bx}, as the contribution from virtual
octets will be a non-resonant signal and the Tevatron Higgs search
would not apply. Due to the fact that the results are reported
only up to Higgs masses of $150 \, {\rm GeV}$ we are only able to
exclude octets up to $75 \rm \, GeV$, which is already disfavoured
by LEP2. If the experimental study of  $h \to \gamma \, \gamma$ is
extended to higher Higgs masses at the Tevatron or LHC, this
signal is likely to be a significant constraint on the model.

We utilize the fact that this signature has been studied for
octetonia in \cite{Kim:2008bx} to demonstrate the potential of
this signal to raise the mass limit on octets. The ratio we are interested in is
that of the
octetonia $\sigma (gg \to O^+) \times BR(O^+ \to \gamma \,
\gamma)$ to the SM rate for  $\sigma (gg \to h) \times BR(h \to
\gamma \, \gamma)$. We take \cite{Kim:2008bx}
\bea
\sigma (gg \to O^+) \times BR(O^+ \to \gamma \, \gamma) \approx \frac{9 \, \pi^3 \, \alpha^2 |\psi(0)|^2}{2 \, M_S \, \hat{s}^2} \delta(1 - m_O^2/\hat{s})
\eea
where $\hat{s}$ is the partonic center of mass energy squared and
$|\psi(0)|$ is the wavefunction at the origin. We have used the
approximation $\rm BR(O^+ \to \gamma \, \gamma) \sim
\alpha^2/\alpha_s^2(2 M_S)$. For the Higgs, we take the
approximation
\bea \sigma (gg \to h) \times BR(h \to \gamma \,
\gamma) \approx \frac{G_F}{\sqrt{2}} \, \frac{M_H^2 \,
\alpha_s^2}{8 \, \pi \, \hat{s}} \, \left(\frac{m_t^4}{M_H^4} \right)
10^{-3} \, \delta(1 - M_H^2/\hat{s}) \eea

Neglecting order one factors the ratio of these two signals scales
as
\bea
R \approx 10^6 \, \frac{\alpha^2}{\alpha_s^2} \, \frac{|\psi(0)|^2}{\hat{s}} \, \left(\frac{M_H^2}{M_S \, m_t^4 \, G_F}\right)
\eea
This ratio must be less than $\sim 35$
\cite{Abazov:2009kq,D05858} for $M_h = (100, 150) \, \rm GeV$ or
$M_{\pm} = (50, 75) \, \rm GeV$. Unless the wavefunction at the
origin was much smaller than its approximate expected value given
by \cite{Kim:2008bx}
\bea
 |\psi(0)|^2 = \frac{N_c^3 \, \alpha_s^3( m_S \, v) \, M_S^3}{8 \, \pi},
\eea
this bound will likely be violated for this entire mass range.
Extending this analysis to higher Higgs masses is expected to
raise the lower mass bound on octet states for this reason. For a
recent comprehensive study of octetonia signals in gamma gamma for
octets from $\sim 200-500$ GeV see \cite{Kim:2008bx}.

\subsection{Flavour constraints reexamined for light scalars}

Flavour constaints on $({\bf8},{\bf2})_{1/2}$ scalars were
examined in some detail in linear MFV\footnote{Where one only
utilizes a linear yukawa coupling for the scalars.} in
\cite{Manohar:2006ga}  when the masses of the octet  scalars were
considered to be $\sim {\rm TeV}$. However, although MFV
suppresses flavour changing effects and ensures the vanishing of
tree level flavour changing neutral currents in linear MFV,  when
one goes beyond leading order in the Yukawa couplings problematic
flavour changing neutral currents are possible
\cite{Gresham:2007ri}. The correct way to examine such flavour
issues is to utilize a nonlinear representation of MFV\footnote{We
thanks J. Zupan for discussions on this point.} such as formulated
in \cite{Feldmann:2008ja,Kagan:2009bn,Feldmann:2009dc} which is
beyond the scope of this work.

We have reexamined the flavour constraints that were examined in
\cite{Manohar:2006ga} in linear MFV for the light octet masses
allowed by EWPD and not ruled out by direct production bounds.
Flavour constraints are largely irrelevant for $|\eta_U|$ once the
far more restrictive constraint from $R_b$ is known. To
quantitatively demonstrate this consider $K^0 - \bar{K}^0$ mixing
for relatively light masses $M_s = 300 \, (400){\rm GeV}$.  We use
the results of  \cite{Manohar:2006ga} for the contribution of the
octets to the wilson coefficient ($C_s$) of the operator
$(V_{td}^\star \, V_{ts})(d_L \, \gamma^\nu \, s_L)(d_L \,
\gamma^\nu \, s_L)$ and use the SM expression of
\cite{Buchalla:1995vs}  for the contribution of this operator to
$K^0 - \bar{K}^0$ mixing and hence $|\epsilon_K|$.  One finds that
the contribution of the octets to  $|\epsilon_K|$ is given by
\bea \Delta |\epsilon_K| = |C_\epsilon \, B_K \, {\rm
Im}[{V_{td}^\star \, V_{ts}}] \,  {\rm Re }[{V_{td}^\star \,
V_{ts}}] \,  C_S|
\eea
Using the measured values $m_K = 497.6 \,
{\rm MeV}$, $f_K = (156.1 \pm 0.8) {\rm MeV}$, $(\Delta M_K)_{exp}
=  3.483 \pm 0.006) \times10^{-12} {\rm MeV}$ one obtains
\bea
C_{\epsilon} = \frac{G_F^2 \, F_K^2 \, m_K \, M_W^2}{6 \, \sqrt{2}
\, \pi^2 \, \Delta M_K} = 3.65 \times 10^4.
\eea
Further, Lattice QCD \cite{Lubicz:2008am} gives the input $B_K(2
\, {\rm GeV}) = 0.54 \pm 0.05$, and using the central values of
fitted values for the CKM parameters $A,\bar{\eta}, \bar{\rho},
\lambda$ from the PDG we find that the shift in $|\epsilon_K|$ is
given by
\bea \Delta |\epsilon_K|  = 1.5 \, (1.6) \times 10^{-12}(
|\eta_U|^2 + 6 \, (3) |\eta_U|^4)
\eea
for $M_s = 300 \, (400){\rm GeV}$. Considering $|\epsilon_K|_{exp}
= (2.229 \pm 0.010)\times 10^{-3}$ while the same values employed
above gives the central value $|\epsilon_K|_{theory} = 1.70\times
10^{-3}$ one can set an upper limit on $|\eta_U|$ from $K^0 -
\bar{K}^0$ mixing by conservatively assigning one tenth of the
discrepency between theory and experiment to the effect of octets.
This gives an upper bound on $|\eta_U|$ of $48\, (56)$ for $M_s =
300 \, (400){\rm GeV}$. The weak mass dependence of the bound
allows one to neglect Kaon mixing constraints for low masses,
compared to $R_b$ constraints on $|\eta_U|$, for light masses $M_s
\ll 1 \, {\rm TeV}$, in linear MFV.

The $B \to X_s \, \gamma$ decay rate constrains the combination
$|\eta_U \, \eta_D|$, in the limit $\eta_U$ is small, and was
calculated in \cite{Manohar:2006ga} . Using their result and the
upper bound on $|\eta_U|$ from $R_b$, we determine the strongest
upper bound on $|\eta_D|$ for light masses by requiring that the
octet contribution to $B \to X_s \, \gamma$ is less than the $\sim
10 \%$ SM theoretical and experimental errors. For $M_{\pm}=
(75,100,200)$ and the corresponding maximum
$|\eta_U|=(0.26,0.27,0.33)$, one obtains an upper bound on
$|\eta_D|$ of $(0.36,0.39,0.50)$. As $|\eta_U|$ decreases, the
upper bound on $|\eta_D|$ is relaxed.

Finally, the electric dipole moment of the neutron constrains the
imaginary part of the $\eta_i$ and using  \cite{Manohar:2006ga} we
find for light masses that ${\rm Im }[\eta_U^\star \,
\eta_D^\star] < 1/10$ for $m_S = 100 \, {\rm GeV}$.

\section{Conclusions}

We have considered the phenomenological constraints of the general
scalar sector that contains one $({\bf1},{\bf2})_{1/2}$ Higgs
doublet and a one $({\bf8},{\bf2})_{1/2}$ colour octet scalar
doublet. To this end we have performed a modern fit in the $\rm
STU$ and $\rm STUVWX$ approaches to EWPD and used these results to
determine the allowed masses for light octets. We have
demonstrated that, somewhat surprisingly, the six parameter fit
formalism is more restrictive for light states due to strong
correlations amongst the fit observables. We find that the octet
doublet masses can be in the $100 \, {\rm GeV}$ range. Such light
octets can significantly effect the discovery strategies for a
light Higgs by modifying the Higgs production mechanism through a
one loop contribution to $gg \to h$ that is not well approximated by a local operator. Octets will also induce a
further effective coupling at one loop between $h$ and $\gamma \,
\gamma$, $Z Z$ and $W^+ W^-$ and would significantly effect Higgs
discovery at LHC \cite{Manohar:2006gz}. Despite this, we have
shown that current production bounds on light octets at LEP2 and
Tevatron do not rule out the low mass region and further studies
for narrow resonances in the dijet invariant mass distribution and
$h \to \gamma \, \gamma$ signal are required. Currently, octets
are another example of physics beyond the SM that can
significantly effect the properties of the Higgs and yet are
otherwise relatively unconstrained experimentally.\footnote{For
further studies of the modification of the properties of the Higgs
through otherwise experimentally elusive new physics see
\cite{Mantry:2007ar,Mantry:2007sj}.} For light octets, one
possible alternate search strategy is to utilize the Higgs $p_T$
distribution \cite{Arnesen:2008fb} to find indirect evidence for
onshell octet scalars that have eluded direct detection.

We have also performed a joint fit for the Higgs and the octets by
varying the Higgs mass oblique corrections at one loop while
allowing the masses of the octets to vary. Doing so we have
demonstrated a mechanism that is quite general in its effect of
giving a positive contribution to the $T$ parameter when an extra
doublet is present and fit to in EWPD. This allows the Higgs and
octet to be simultaneously heavy and the Higgs can be as massive
as its unitarity bound.  For the parameter space where the Higgs
mass is raised, $h$ decaying to pairs of octets is kinematically
suppressed. The search strategy for the heavy Higgs remains
substantially the same with difficulties in constructing a mass
peak due to the width of the Higgs resonance and large irreducible
backgrounds to SM processes producing $W^+ \, W^-$ decays such as
from $\bar{t} \, t$, and large $W j$ backgrounds.
Likewise very heavy octets are also very broad resonances for
large masses and are difficult to discover at hadron colliders
with decays to $\bar{t} t$ dominating, and  large SM backgrounds.
Further dedicated studies of LHC phenomenology of this scenario are warranted, as
are further dedicated studies to attempt to raise the lower mass
bounds on octet scalar doublets.

\section*{Acknowledgements}
We sincerely thank B Batell for extensive collaboration during this work.

We also thank Mark Wise, Maxim Pospelov and Aneesh Manohar for comments on the manuscript. 
We  thank Michael Luke for many helpful discussions and Jens Erler for communication on the EWPD fits.
We are particularly grateful to Aneesh Manohar for an extensive debate on the precise nature of the 
new $\rm \, SU(2)_C$ constraint on the potential, which he won.

This work was partially supported by funds from the Natural
Sciences and Engineering Research Council (NSERC) of Canada.
Research at the Perimeter Institute is supported in part by the
Government of Canada through NSERC and by the Province of Ontario
through MRI.
\vfill\eject

\pagebreak

\appendix

\section{EWPD fit}

The data and theory predictions used in constructing the fit are given in Table 2.

\begin{table}[h]
\begin{center}
\begin{tabular}{|c||c||c|} \hline
Observable &  Data Used & Theory Prediction  \\
\hline \hline
$M_W$ [GeV] & 80.428 $\pm$ 0.039 & 80.380 $\pm$ 0.015  \\
 & 80.376 $\pm$ 0.033 & 80.380 $\pm$ 0.015  \\
$M_Z$ [GeV] & 91.1876 $\pm$ 0.0021 & 91.1874 $\pm$ 0.0021  \\
$\Gamma_Z$ [GeV] & 2.4952 $\pm$ 0.0023 & 2.4954 $\pm$ 0.0009 \\
$\Gamma_{had}$ [GeV] & 1.7444 $\pm$ 0.0020 & 1.7419 $\pm$ 0.0009  \\
$\Gamma_{inv}$ [MeV  & 499.0 $\pm$ 1.5 & 501.68 $\pm$ 0.07  \\
$\Gamma_{l^+ l^-}$ [MeV] & 83.984 $\pm$ 0.086 & 84.002 $\pm$ 0.016  \\
$\sigma_{had}$ [nb] & 41.541 $\pm$ 0.037 & 41.483 $\pm$ 0.008 \\
$R_e$ & 20.804 $\pm$ 0.050 & 20.736 $\pm$ 0.010 \\
$R_\mu$ & 20.785 $\pm$ 0.033 & 20.736 $\pm$ 0.010 \\
$R_\tau$ & 20.764 $\pm$ 0.045 & 20.736 $\pm$ 0.010 \\
$R_b$ & 0.21629 $\pm$ 0.00066 & 0.21578 $\pm$ 0.00005 \\
$R_c$ & 0.1721 $\pm$ 0.0030 & 0.17224 $\pm$ 0.00003  \\
$A_{FB}^e$ & 0.0145 $\pm$ 0.0025 & 0.01627 $\pm$ 0.00023  \\
$A_{FB}^\mu$ & 0.0169 $\pm$ 0.0013 & 0.01627 $\pm$ 0.00023  \\
$A_{FB}^\tau$ & 0.0188 $\pm$ 0.0017 & 0.01627 $\pm$ 0.00023 \\
$A_{FB}^b$ & 0.0992 $\pm$ 0.0016 & 0.1033 $\pm$ 0.0007  \\
$A_{FB}^c$ & 0.0707 $\pm$ 0.0035 & 0.0738 $\pm$ 0.0006  \\
$\bar{s}_l^2(A_{FB}^q)$ & 0.2316 $\pm$ 0.0018 & 0.2315 $\pm$ 0.0001 \\
$A_e$ & 0.15138 $\pm$ 0.00216 & 0.1473 $\pm$ 0.0010 \\
& 0.1544 $\pm$ 0.0060 & 0.1473 $\pm$ 0.0010 \\
& 0.1498 $\pm$ 0.0049 & 0.1473 $\pm$ 0.0010\\
$A_\mu$ & 0.142 $\pm$ 0.015 & 0.1473 $\pm$ 0.0010 \\
$A_\tau$ & 0.136 $\pm$ 0.015 & 0.1473 $\pm$ 0.0010 \\
& 0.1439 $\pm$ 0.0043 & 0.1473 $\pm$ 0.0010 \\
$A_b$ & 0.923 $\pm$ 0.020 & 0.9347 $\pm$ 0.0001 \\
$A_c$ & 0.670 $\pm$ 0.027 & 0.6679 $\pm$ 0.0004  \\
$g_L^2$ & 0.3010 $\pm$ 0.0015 & 0.3039 $\pm$ 0.0002 \\
$g_R^2$ & 0.0308 $\pm$ 0.0011 & 0.03000 $\pm$ 0.00003  \\
$g_V^{\nu e}$ & -0.040 $\pm$ 0.015 & -0.0397 $\pm$ 0.0003  \\
$g_A^{\nu e}$ & -0.507 $\pm$ 0.014 & -0.5064 $\pm$ 0.0001  \\
$Qw(Cs)$ & -73.16 $\pm$ 0.35 & -73.16 $\pm$ 0.03  \\
$Qw(Tl)$ & -116.4 $\pm$ 3.6 & -116.8 $\pm$ 0.04  \\
$\Gamma_W$ [GeV] & 2.141 $\pm$ 0.041 &  2.0902 $\pm$ 0.0009 \\
\hline
\end{tabular} \label{data}
\caption{Observables used in fit to oblique parameters.}
\end{center}
\end{table}

The numbers we use for the theory predictions are based on the $2008$ PDG results of a global fit to the EWPD.
The input values used in the theory predictions are
\ba
M_Z &=& 91.1876 \pm 0.0021 {\rm GeV},  \quad \quad M_H = 96^{+29}_{-24} {\rm GeV}, \nn \\
m_t &=& 173.1 \pm 1.4 {\rm GeV},  \quad \quad \, \,  \, \, \,  \alpha_s(M_Z) = 0.1217 \pm 0.0017 {\rm GeV}, \\
\hat{\alpha}(M_Z)^{-1} &=& 127.909 \pm 0.0019, \quad \quad  \, \, \, \, \, \Delta \, \alpha_{had}^{(5)} \approx 0.02799 \pm 0.00014.\nn
\ea

The definitions of the oblique corrections we use are
\begin{eqnarray}
 \frac{\alpha S}{4 s_W^2 \, c_W^2} &=& \left[\frac{\delta \Pi_{ZZ}(M_Z^2) - \delta \Pi_{ZZ}(0)}{M_Z^2} \right]
 - \frac{(c_W^2 - s_W^2)}{s_W \, c_W}  \delta \Pi'_{Z\, \gamma}(0)  -  \delta \Pi'_{\gamma \, \gamma}(0), \nn \\
 \alpha T &=& \frac{\delta \Pi_{WW}(0)}{M_W^2} - \frac{\delta \Pi_{ZZ}(0)}{M_Z^2}, \nn \\
 \frac{\alpha U}{4 s_W^2} &=& \left[\frac{\delta \Pi_{WW}(M_W^2) - \delta \Pi_{WW}(0)}{M_W^2} \right]
 - c_W^2 \left[\frac{\delta \Pi_{ZZ}(M_Z^2) - \delta \Pi_{ZZ}(0)}{M_Z^2} \right] \nn \\
&\,& - s_W^2 \, \delta \Pi'_{\gamma\, \gamma}(0)  -  2 \, s_W \, c_W \, \delta \Pi'_{Z \, \gamma}(0),\\
  \alpha V &=&  \delta \Pi'_{Z Z}(M_Z^2) - \left[\frac{\delta \Pi_{ZZ}(M_Z^2) - \delta \Pi_{ZZ}(0)}{M_Z^2} \right], \nn \\
  \alpha W &=&  \delta \Pi'_{WW}(M_W^2) - \left[\frac{\delta \Pi_{WW}(M_W^2) - \delta \Pi_{WW}(0)}{M_W^2} \right], \nn \\
  \alpha X &=&  - s_W \, c_W \left[ \frac{\delta \Pi_{Z \, \gamma}(M_Z^2)}{M_Z^2} -  \delta \Pi'_{Z \, \gamma}(0) \right] \nn
\end{eqnarray}

The self energies to determine these results are given by the following in terms of PV functions that match the definitions in \cite{Wells:2005vk} and are

\begin{eqnarray}
16 \, \pi^2 \, \mu^{4-n} \, \int \, \frac{d^n \, q}{i \, (2 \, \pi)^n} \, \frac{1}{q^2 - m^2 + i \, \epsilon} &=& A_0(m^2)  \\
16 \, \pi^2 \, \mu^{4-n} \, \int \, \frac{d^n \, q}{i \, (2 \, \pi)^n} \, \frac{1}{[q^2 - m_1^2 + i \, \epsilon]\,[(q-p)^2 - m_2^2 + i \, \epsilon]} &=& B_0(p^2,m_1^2,m_2^2) \nn \\
16 \, \pi^2 \, \mu^{4-n} \, \int \, \frac{d^n \, q}{i \, (2 \, \pi)^n} \, \frac{q_\mu}{[q^2 - m_1^2 + i \, \epsilon]\,[(q-p)^2 - m_2^2 + i \, \epsilon]} &=& p_\mu \, B_1(p^2,m_1^2,m_2^2) \nn \\
16 \, \pi^2 \, \mu^{4-n} \, \int \, \frac{d^n \, q}{i \, (2 \, \pi)^n} \, \frac{q_\mu \, q_\nu}{[q^2 - m_1^2 + i \, \epsilon]\,[(q-p)^2 - m_2^2 + i \, \epsilon]} &=& p_\mu \, p_\nu \,  B_{21}(p^2,m_1^2,m_2^2), \nn \\
&\,& + g_{\mu \, \nu} \,  B_{22}(p^2,m_1^2,m_2^2) \nn
\end{eqnarray}

Our results are
\begin{eqnarray}
\delta \Pi_{WW}(p^2) &=& \frac{g_1^2}{2 \pi^2}\Big[   B_{22}(p^2,M_I^2,M_+^2) \nn
+  B_{22}(p^2,M_R^2,M_+^2)\\ &-& \frac{1}{2} A_0(M_+^2)-\frac{1}{4} A_0(M_R^2)-\frac{1}{4} A_0(M_I^2)\Big] \\
\delta \Pi_{ZZ}(p^2) &=& \frac{g_1^2 }{2 \pi^2 c_W^2}\Big[
(1-2s_W^2)^2 \left( B_{22}(p^2,M_+^2,M_+^2) -\frac{1}{2} A_0(M_+^2)\right) \nn \\
&+&  B_{22}(p^2,M_R^2,M_I^2 )-\frac{1}{4} A_0(M_R^2)-\frac{1}{4} A_0(M_I^2)\Big] \\
\delta \Pi_{\gamma\gamma}(p^2) &=& \frac{2 e^2 }{ \pi^2 }\Big[
 B_{22}(p^2,M_+^2,M_+^2) -\frac{1}{2} A_0(M_+^2) \Big]\\
\delta \Pi_{\gamma Z}(p^2) &=& \frac{e g_1(1-2s_W^2)}{ \pi^2 c_W }\Big[
 B_{22}(p^2,M_+^2,M_+^2) -\frac{1}{2} A_0(M_+^2) \Big]
\end{eqnarray}

For $p^2=0$ these expressions become
\begin{eqnarray}
\delta \Pi_{WW}(0) &=& \frac{g_1^2 }{8 \pi^2} \left( \frac{1}{2} f(M_+,M_R) +  \frac{1}{2} f(M_+,M_I) \right) \\
\delta \Pi_{ZZ}(0) &=& \frac{g_1^2}{8 \pi^2 c_W^2} \left( \frac{1}{2} f(M_R,M_I) \right)
\end{eqnarray}
where
\begin{equation}
 f(m_1,m_2)=m_1^2 +m_2^2 -\frac{2 m_1^2 m_2^2}{m_1^2-m_2^2}\log{ \frac{m_1^2}{m_2^2} }
\end{equation}

The derivatives of the vacuum polarizations are
\begin{eqnarray}
\delta \Pi'_{\gamma\gamma}(0)&=& - \frac{e^2}{6 \pi^2} B_0(0,M_+^2,M_+^2) \\
\delta \Pi'_{\gamma Z}(0) &=& -\frac{e g_1(1-2s_W^2)}{12 \pi^2 c_W } B_0(0,M_+^2,M_+^2)
\end{eqnarray}

\section{Renormalization}

We use dim reg in $d = 4 - 2 \, \epsilon$ dimensions. We introduce
wavefunction renormalization and mass renormalization constants
for the octet fields as usual \bea S_i =
\frac{S_i^{(0)}}{\sqrt{Z_{i}}}, \quad \quad M_i =
\frac{M_i^{(0)}}{\sqrt{Z_{Mi}}}. \eea However, in choosing
renormalization conditions, we note that to define the masses and
the mass splittings one cannot use $\rm{\overline{MS}}$, as in
$\rm{\overline{MS}}$ the mass is defined to have only the
divergence subtracted from the bare mass. The resulting
renormalized mass in $\rm{\overline{MS}}$ is not shifted by the
finite components of the loop corrections that we have determined.
The renormalization prescription we use is the zero-momentum
subtraction scheme \cite{Bogoliubov:1957gp},  where we require
that the self energy and its derivative with respect to external
momentum, $p^2$, vanishes at $p^2 \to 0$. Note that the second
derivative term in the Taylor expansion of the self energy does
not contribute until two loop order and therefore can be neglected
here. The counter terms in the lagrangian are given by \bea \sum_i
\left[(Z_i - 1) (\partial^\mu \, S_i \, \partial_\mu S_i) - (Z_i \
Z_{Mi} - 1) M_i^2 \, S_i^2\right]. \eea With this prescription the
wavefunction renormalization and the mass counterterm are of the
form \bea
Z_i &=& 1- \frac{a}{\epsilon}-\frac{d \, \Sigma_i(p^2)}{d \, p^2} \mid_{p^2 \rightarrow 0} \nn \\
Z_{Mi} &=& 1 +\frac{b}{\epsilon}+  \Sigma_i(p^2) \mid_{p^2 \rightarrow 0} + (1-Z_i)
\eea
where $a,b$ are the coefficients of the $p^2,M^2$ dependent one loop divergences respectively and the $\Sigma_i$ are the finite terms of the one loop self energy.

Using this scheme and the divergence properties of the PV functions, the wavefunction renormalization factors are determined to be
\bea
Z_I  &=& 1 - \frac{y_t^2 \, |\eta_U|^2 + y_b^2 \, |\eta_D|^2}{64 \, \pi^2  \, \epsilon} + \frac{g_1^2}{32 \, \pi^2 \, \epsilon}\left[1 + \frac{1}{2 \, c_W^2}\right]  + \frac{y_t^2 |\eta_U|^2 \, \log \left[\frac{m_t^2}{\mu^2} \right] + y_b^2 |\eta_D|^2 \, \log \left[\frac{m_b^2}{\mu^2} \right] }{32 \, \pi^2} \nn \\
&\,&+ \frac{y_t^2 \, {\rm Im}[\eta_U]^2 + y_b^2 \, {\rm Im}[\eta_D]^2}{48 \, \pi^2}   + \frac{g_1^2}{16 \, \pi^2} \, \left[b_0(0,M_W^2,m_{\pm}^2) + \frac{b_0(0,M_Z^2,M_{R}^2)}{2 \,c_W^2} \right] \nn \\
Z_R  &=& 1 - \frac{y_t^2 \, |\eta_U|^2 + y_b^2 \, |\eta_D|^2}{64 \, \pi^2  \, \epsilon} + \frac{g_1^2}{32 \, \pi^2 \, \epsilon}\left[1 + \frac{1}{2 \, c_W^2}\right]
 + \frac{y_t^2 |\eta_U|^2 \, \log \left[\frac{m_t^2}{\mu^2} \right] + y_b^2 |\eta_D|^2 \, \log \left[\frac{m_b^2}{\mu^2} \right] }{32 \, \pi^2} \nn \\
&\,& + \frac{y_t^2 \, {\rm Re}[\eta_U]^2 + y_b^2 \, {\rm Re}[\eta_D]^2}{48 \, \pi^2}
+ \frac{g_1^2}{16 \, \pi^2} \, \left[b_0(0,M_W^2,M_{\pm}^2) + \frac{b_0(0,M_Z^2,M_{I}^2)}{2 \,c_W^2} \right] \nn \\
Z_{\pm} &=&  1 - \frac{y_t^2 \, |\eta_U|^2 + y_b^2 \, |\eta_D|^2}{64 \, \pi^2 \, \epsilon} + \frac{g_1^2}{32 \, \pi^2 \, \epsilon}\left[1 + \frac{(1 - 2 \, s_W^2)^2}{2 \, c_W^2} + 2 \, s_W^2 \right]  \nn \\
&+&  \frac{g_1^2}{32 \, \pi^2} \left[b_0(0,M_W^2,M_{I}^2)+b_0(0,M_W^2,M_{R}^2)
 + \frac{(1-2 s_W^2)^2 b_0(0,M_Z^2,M_{\pm}^2)}{c_W^2} -4 s_W^2 \left( \log \left[\frac{M_{\pm}^2}{\mu^2}\right] -1\right)  \right] \nn \\
&-& \frac{(y_b^2 | \eta_D|^2 + y_t^2 | \eta_U|^2)}{32 \, \pi^2} \, b_0(0,m_b^2,m_t^2)
\eea

Using these results the mass renormalization factors are determined to be
\bea
Z_{MI} &=& (2 - Z_I)  - \frac{v^2}{32\, \pi^2  \, M_I^2} \left[ y_t^4 \, ({\rm Re}[\eta_U]^2 + 3 \, {\rm Im}[\eta_U]^2) \left(\frac{1}{2\, \epsilon}-\log \left[\frac{m_t^2}{\mu^2}\right]\right) \right. \\
&\,&\, \, \, \left.+ y_b^4 \, ({\rm Re}[\eta_D]^2 + 3 \, {\rm Im}[\eta_D]^2) \left(\frac{1}{2\, \epsilon}-\log \left[\frac{m_b^2}{\mu^2}\right]\right) \right] -\frac{v^2 \left(y_t^4 |\eta_U|^2+y_b^4 |\eta_D|^2\right)}{32 \pi^2 \, M_I^2} \nn  \\
&\,&\, \, \,  + \frac{g_1^2}{64 \,\pi^2 \, M_{I}^2 \, \epsilon } \left[ 3\,M_W^2 - M_\pm^2 + \frac{(3\, M_Z^2 - M_R^2)}{2 \, c^2 } \right]  \nn \\
&\,&\, \, \, + \frac{g_1^2}{32 \, \pi^2 \,M_I^2} \left[ (M_W^2-2 M_\pm^2)\, b_0[0,M_W,M_\pm] +\frac{(M_Z^2-2 \, M_R^2) \, b_0[0,M_Z,M_R]}{2 c_W^2} \right. \nn \\
&\,&\, \, \, + M_\pm^2 \left( 1- \log \left[\frac{M_\pm^2}{\mu^2}\right] \right) +\, M_W^2\left(1-2\, \log \left[\frac{M_W^2}{\mu^2}\right]\right)+ \frac{M_Z^2}{2 \, c_W^2} \left(1-2\,\log \left[\frac{M_Z^2}{\mu^2}\right]\right) \nn \\
&\,&\, \, \ \left. +\frac{M_R^2}{2\,c_W^2} \left(1-\log \left[\frac{M_R^2}{\mu^2}\right]\right) \, \right]  \nn \\
Z_{MR} &=& Z_{MI} \mid_{M_R^2 \rightarrow M_I^2, Z_I \rightarrow Z_R, {\rm Re}\leftrightarrow  {\rm Im}} \nn \\
Z_{M\pm} &=&(2 - Z_{\pm})  - \frac{v^2\,y_b^4 \, |\eta_D|^2}{64 \, \pi^2 \, M_{\pm}^2} \left[\frac{1}{\epsilon}+b_0[0,m_b,m_t] - \log \left[\frac{m_b^2}{\mu^2}\right]+1\right] \nn \\
&\,&\, \, \, - \frac{v^2\,y_t^4 \, |\eta_U|^2}{64 \, \pi^2 \, M_{\pm}^2} \left[\frac{1}{\epsilon}+b_0[0,m_b,m_t] - \log \left[\frac{m_t^2}{\mu^2}\right]+1\right] \nn \\
&\,&\, \, \, - \frac{y_b^2 \, y_t^2 v^2 }{64 \, \pi^2\, M_{\pm}^2} \left[  |\eta_D|^2 \left(\frac{1}{\epsilon}-\log \left[\frac{m_t^2}{\mu^2}\right] +1 +b_0[0,m_b,m_t] \right) \right. \nn \\
&\,&\, \, \,  \left.  +  |\eta_U|^2 \left(\frac{1}{\epsilon}- \log \left[\frac{m_b^2}{\mu^2}\right] +1 +b_0[0,m_b,m_t] \right)   -  (\eta_D \, \eta_U +  \, \eta^\star_D \, \eta^\star_U) \left(\frac{1}{\epsilon}+2\,b_0[0,m_b,m_t]\right) \right]    \nn \\
&\,&\, \, \, + \frac{g_1^2}{32 \,\pi^2 \, \epsilon } \left[\frac{6 \, M_W^2 - M_R^2 - M_I^2}{4 \, M_{\pm}^2} + \frac{(1 - 2 s_W^2)^2}{4 \, c_W^2} \frac{( 3 \, M_Z^2 - M_{\pm}^2)}{M_{\pm}^2} - s_W^2\right] \nn \\
&\,&\, \, \, + \frac{g_1^2}{64 \,\pi^2 \,M_\pm^2} \Big[ (M_W^2-2\,M_I^2)\,b_0[0,M_W,M_I]+(M_W^2-2\,M_R^2)\,b_0[0,M_W,M_R]  \nn \\
&\,&\, \, \, +(M_Z^2-2\,M_\pm^2) \frac{(1-2s_W^2)^2}{c_W^2} b_0[0,M_Z,M_\pm] +M_I^2 \left(1-\log \left[ \frac{M_I^2}{\mu^2}\right] \right) \nn \\
&\,&\, \, \, +M_R^2 \left(1-\log \left[ \frac{M_R^2}{\mu^2}\right] \right)+2 \,M_W^2 \left(1-2 \log \left[ \frac{M_W^2}{\mu^2}\right] \right) \nn \\
&\,&\, \, \, + \frac{M_Z^2 \,(1-2s_W^2)^2}{c_W^2}  \left(1-2 \log \left[ \frac{M_Z^2}{\mu^2}\right] \right) +M_\pm^2 \frac{8 s_W^4-8 s_W^2+1}{c_W^2}\left(1- \log \left[ \frac{M_{\pm}^2}{\mu^2}\right] \right) \Big]
\eea

The remaining renormalization is for the mixing operator $S_R \, S_I$ which is renormalized as
usual by introducing a further counter term to subtract the only divergences of composite operators as in $\rm{\overline{MS}}$
 \bea
 \frac{\sqrt{Z_I} \, \sqrt{Z_R} \, (v^2 \,S_R \, S_I)}{Z_{RI}}
\eea
where
\bea
Z_{RI} = 1+\frac{Z_R -1}{2} +  \frac{Z_I -1}{2} + \frac{y_t^4 \,  {\rm Re}[\eta_U] \,  {\rm Im}[\eta_U] - y_b^4 \,  {\rm Re}[\eta_D] \,  {\rm Im}[\eta_D]}{32 \, \pi^2 \ \epsilon}
\eea

\section{Mixing of $S_R$ and $S_I$}

For completeness in examining one loop effects we determine the
mixing between $S_R$ and $S_I$. The mass matrix is given by \bea
M_{RI} = \left(\begin{array}{cc} M_I^2 + \delta \langle T\{S^I \, S^I\} \rangle_G + \delta \langle T\{S^I \, S^I\} \rangle_Y & \delta \langle T\{S^R \, S^I\} \rangle_Y \\
 \delta \langle T\{S^R \, S^I\} \rangle_Y & M_R^2 + \delta \langle T\{S^R \, S^R\} \rangle_G + \delta \langle T\{S^R \, S^R\} \rangle_Y \end{array} \right). \nn \\
\eea
We diagonalize the mass matrix by introducing a mixing angle and rotating the $S_R,S_I$ fields to a diagonal mass basis
$S_R',S_I'$ via
\bea
 \left(\begin{array}{c} S_I \\
S_R \end{array} \right)  =  \left(\begin{array}{cc} \cos(\theta) & \sin(\theta) \\
- \sin(\theta)  & \cos(\theta) \end{array} \right)  \left(\begin{array}{c} S_I '\\
S_R' \end{array} \right).
\eea
The mixing angle is given by
\bea
\sin(\theta) = \frac{|y_t^4 \, B_0^\star(p^2,m_t^2,m_t^2){\rm Re}(\eta_U) \,{\rm Im}(\eta_U) - y_b^4 \,B_0^\star(p^2,m_b^2,m_b^2){\rm Re}(\eta_D) \,{\rm Im}(\eta_D)|}{8 \, \pi^2 \, \lambda_2}
\eea
where $B_0^\star$ is the usual PV function with the divergence subtracted given by
\bea
 B_0^\star(p^2,m_i^2,m_i^2) =  -2 + \log \left(\frac{m_i^2}{\mu^2}\right) - \beta \, \log \left(\frac{1+ \beta}{1 - \beta} \right)
\eea
where $\beta = \sqrt{1 - 4 m_i^2/p^2}$, which would be the velocity of the scalar produced in the CM frame which was subsequently to mix into another state with mass $m_j$. We take $p^2 = m_s^2$ as the mass splittings are a small  perturbation in a radiatively induced mixing. If we take $\mu \simeq 1 \, {\rm TeV}$ as the scale at which we impose exact $\rm SU(2_C)$ on our scalar potential,  this gives a mixing angle
\bea
\sin(\theta) \simeq 0.04 \,  \frac{|{\rm Re}(\eta_U)| \,|{\rm Im}(\eta_U)|}{\lambda_2},
\eea
which depends weakly on the value of $m_s$  as the numerical coefficient changes by $25 \%$ for $m_s$ varying between $0.01 - 300$ $\rm GeV$. This mixing angle, if non zero, will effect the production cross section of the
$S_I,S_R$ states at LHC and Tevatron, and introduce mixing between the octetonia states discussed in \cite{Kim:2008bx}.

\end{document}